\documentclass[12pt]{article}
\usepackage{amsfonts}
\usepackage{amsmath}
\usepackage{amssymb}
\usepackage{epsfig}
\usepackage{color}
\usepackage{float}
\usepackage{tabularx}
\usepackage{ulem}
\usepackage{cancel}

\definecolor{darkblue}{rgb}{0,0,.5}

\begin{document}

\begin{flushright}
November 2025\\
\end{flushright}
\vspace{5mm}
\begin{center}
\Large {\bf Dark Energy and the Time Dependence of Fundamental Particle Constants}\\
\mbox{ }\\
\normalsize
\vspace{3.5cm}
{\bf Bodo Lampe} \\              
\vspace{0.3cm}
II. Institut f\"ur theoretische Physik der Universit\"at Hamburg \\
Luruper Chaussee 149, 22761 Hamburg, Germany \\
%e-mail: Lampe.Bodo@web.de \\   

\vspace{4.0cm}

%\begin{figure}[H]
%%%%%[H] in verbindung mit package float
%\begin{center}
%%%%\epsfig{file=fig1philo.eps,height=5.4cm}
%\epsfig{file=Bodolampe2014a.eps,height=6.0cm}
%%%%\epsfig{file=Bodolampe2008.eps,height=6.0cm}
%\end{center}
%\end{figure}

%\vspace{0.8cm}
%\vspace{3.0cm}
{\bf Abstract}\\
\end{center} 
%abstract vage halten?
The cosmic time dependencies of $G$, $\alpha$, $\hbar$ and of Standard Model parameters like the Higgs vev and elementary particle masses are studied in the framework of a new dark energy interpretation. Due to the associated time variation of rulers, many effects turn out to be invisible. However, a rather large time dependence is claimed to arise in association with dark energy measurements, and smaller ones in connection with the Standard Model. %Except for alpha, all of the constants mentioned are dimensionful. Therefore, due to the associated time dependence of rulers, most of the effects cannot be measured. In contrast, the predicted variation of alpha is small, but measurable. 
%\thispagestyle{empty} %dies wuerde die erste Seite ohne Seitenzahl
%wegl in arxiv(The dark energy equation of state and a formula on the full size of the universe are derived in the appendix.)
%statt R(n) lieber die reellen Zahlen  \mathbb{C}
%tensor ist \otimes
%\frac{1}{2}
%statt slash bar f soll man in den nichtrelativistischen 
%Ausruecke f kreuz verwenden! \dagger

\newpage

\normalsize
%\large
%DIES GIBT GRoeSSERE SCHRIFT

%\section{nicht vergessen} 

%--------------------------------------------------------

%\newpage
%\vspace{3.0cm}

%\vspace{0.3cm}
%\begin{flushright}
%\emph{Der Kampf der Vernunft besteht darin,}\\
%\emph{dasjenige, was der Verstand fixiert hat,}\\ 
%\emph{zu \"uberwinden.}\\
%{\sc G. W. F. Hegel}
%\end{flushright}

%\color{blue}{$\Lambda c^/3$}
%\color{red}{$\Lambda c^/3$}
%\color{green}{$\Lambda c^/3$}

\begin{center}
{\bf I. Introduction}
\end{center}

Dirac was one of the first to suggest that fundamental physical constants may vary in time due to the expansion of the universe\cite{dirac}. Dirac concentrated on Newton's constant G, but since then a time dependence of $c$, $\alpha$, $\hbar$ and so on has been considered possible as well (\cite{uzan}-\cite{shaw1}). 

From the 21st century perspective it is clear that if fundamental constants are time dependent in this way, the observed dark energy effect must have to do with it, because dark energy dominates the expansion of the universe. 

In the course of the present work {\bf time dependencies} will therefore be partly reduced to a {\bf (dark) energy dependence} of physical quantities and constants. This energy dependence is completely separate from and not to be confused with the usual energy dependence from the renormalization group.

The framework of the article will be the ordinary FLRW cosmology with a scale factor a(t) and a spatial curvature k, the latter assumed to be tiny (in accordance with observations). Furthermore, the so-called 'cosmic coordinate system' will be used, i.e. cosmic time t and proper distances r as parameters. This will prove to be optimal for the presentation. 

It is well known that the fundamental spacetime constants $c$, $\hbar$ and G can be used to define the Planck length, time and mass L, T and M which describe the basic properties of space[m], time[s] and matter[kg]
%Fundamental length, time and mass scales can be given in terms of these quantities as 
\begin{equation} 
L(t)=\sqrt{\frac{\hbar (t) G(t)}{c^3}} \qquad \quad
T(t)=\sqrt{\frac{\hbar (t) G(t)}{ c^5}} \qquad \quad
M(t)=\sqrt{\frac{\hbar (t) c}{ G(t)}} 
\label{tm33b}
\end{equation}
One may invert these relations to obtain
\begin{eqnarray} 
c =\frac{L(t)}{T(t)} \qquad \quad
\hbar (t) =E(t) T(t) \qquad \quad
\kappa (t) = \frac{L(t)}{E(t)} 
\label{tm33bx}
%c &=&\frac{L(t)}{T(t)} \label{tm33bxc}\\
%\hbar (t) &=&E(t) T(t) \label{tm33bxh} \\
%\kappa (t) &=& \frac{L(t)}{E(t)} \label{tm33bx}
\end{eqnarray}
where $E=Mc^2$ is the Planck energy and $\kappa =G/c^4$ the Einstein constant.

A time dependence of these quantities has been anticipated here. $t=0$ is taken to be the present, so we have the present day values $L_0=L(0)=$ Planck length, $T_0=T(0)=$ Planck time and $E_0=E(0)=$ Planck energy. Numerical values are
\begin{eqnarray} 
%L_0 = 1.6\times 10^{-35} m\qquad \quad
%E_0/c^2 = 2.2\times 10^{-8} kg
L_0 = 1.6\times 10^{-35} m\qquad \quad
M_0 = 2.2\times 10^{-8} kg\qquad \quad
T_0 = 5.4 \times 10^{-44} s
\label{tm3num}
\end{eqnarray}

No time dependence of c is indicated, because in the present model there is none - at least if one uses the above mentioned cosmic coordinates t and r, in which case the FLRW solution of the Einstein equations has the line element 
\begin{equation} 
ds^2=-c^2 dt^2+a(t)^2 \,[\frac{dr^2}{1-kr^2}+r^2 d\Omega^2\,]   
\label{tmas33b}
\end{equation}
with a constant i.e. time-independent speed of light.
%\footnote{If one would be willing to modify the FLRW equations by a time-dependent c, one would be in conflict to the framework developed below, cf \cite{tetrons}.}
%Note, the often used 'comoving' coordinate is given by $r_c:=a(t)r$.
%and this can be used to define the FLRW expansion parameter a. 

c being constant, one only needs to consider time dependencies of G and h. 

Equivalently, since one has T(t)=L(t)/c one only needs to consider time dependencies of the Planck length L(t) and Planck energy E(t). 

Rewriting eq. (\ref{tm33bx}) as
\begin{eqnarray} 
\hbar(t)c&=&E(t)L(t) \label{tmeins}\\
G(t)&=&c^4L(t)/E(t)  \label{tmzwei} 
\end{eqnarray}
one sees that there are 2 really fundamental time dependencies to be considered:\\
\colorbox{yellow}{-L(t)=the time dependence of the fundamental measure of space}
\\ 
\colorbox{yellow}{-E(t)=the time dependence of the 'physically active' quantities - the 'quantities} 
\\
\colorbox{yellow}{of motion', as Isaac Newton called them.}

%\begin{addmargin}[25pt]{0pt} 
%{\small Remark: The time dependence of elementary particle couplings $\alpha$, $G_F$ and so on is a %different story. It will be treated in section V and will boil down to determine the time dependence %of one other quantity\\ 
%{\bf -J(t)=the time dependence of the 'internal exchange energy'} to be defined in section V.}
%\end{addmargin}

The cosmic time dependence of elementary particle couplings like the fine structure constant $\alpha$, the elementary particle masses and so on is a different story. It will be treated in section V and will boil down to determine the time dependence of one other quantity:\\ 
\colorbox{yellow}{
-J(t)=the time dependence of the 'internal exchange energy' to be defined in section V.
}

{\bf  In order to determine L(t), E(t) and J(t), in sections II, III and V equations (\ref{eq551x}), (\ref{ab222}) and (\ref{ab1222}) will be introduced.}

\begin{center}
{\bf II. Measure-of-Space Equation}
\end{center}

For L(t) the following equation is suggested:
\begin{eqnarray}
\ddot{L}=-\frac{4\pi}{3} G\rho L-\omega^2 (L-L_s) + \xcancel{\frac{\Lambda}{3} c^2 L}
\label{eq551x}
\end{eqnarray}
%d2L/dt2=-G*L*rho/3-w2*(L-Ls)    (3)

\begin{figure}
\begin{center}
\epsfig{file=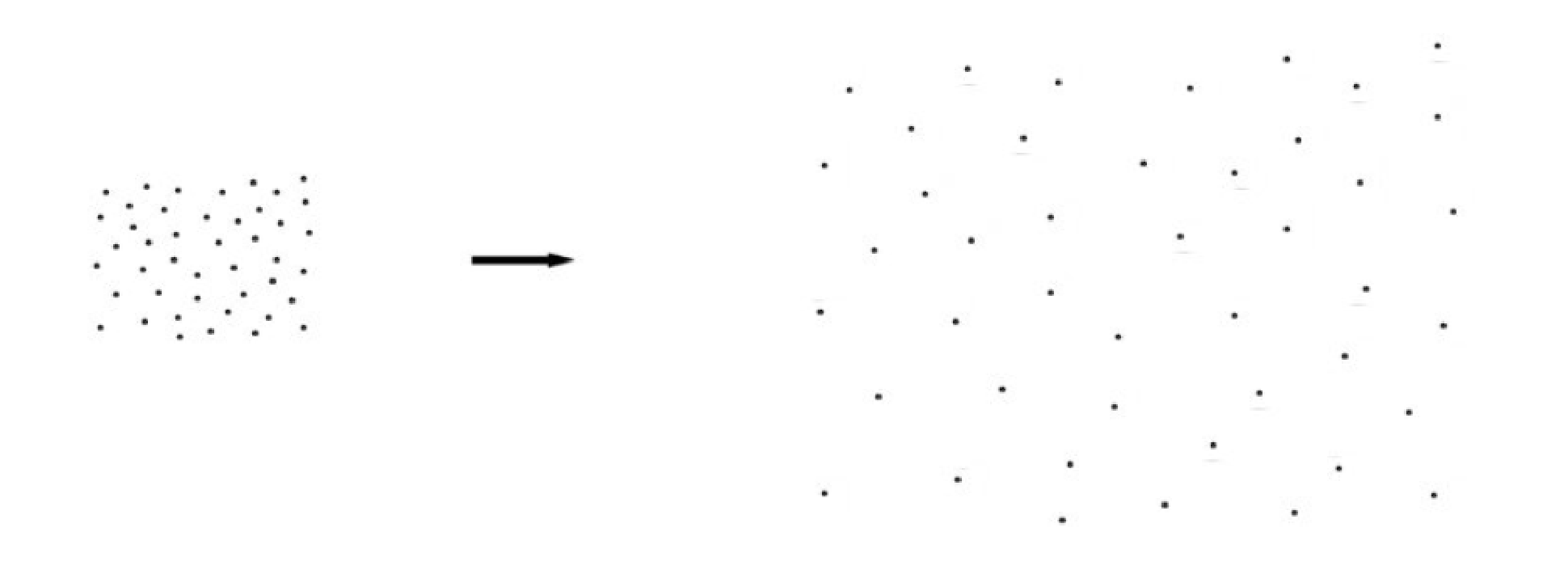,height=5.4cm}
%\bigskip
\caption{Schematic depiction of the universe as an elastic substrate made of tiny constituents at 2 times $t_1 < t_2$. In the full tetron model, the 'constituents' are actually tetrahedrons composed out of 'tetrons' and extending into 3 extra dimensions, cf. Fig. 3. However, in physical space, the tetrahedrons are pointlike objects, and since particle physics interactions do not play a role in this part of the paper, one can forget about the internal structure.}
\nonumber
%\label{figac1} 
\end{center}
\end{figure}

The idea behind this is that the universe is an elastic substrate which consists of elementary constituents called tetrons, and the bond length of these constituents is given by the Planck length L(t), while the Planck energy E(t) measures the binding energy of every 2 bound constituents. Furthermore, the universe is carrying out an extremely low frequency breathing vibration around $L_s$, and all the tetron bonds on average are vibrating in accord with the universe. Details on this microscopic model can be found in \cite{tetrons,lhiggs}. It is to be noted that the tetrons are invisible to us. All (ordinary and dark) matter particles and radiation we know are quasi-particles/wave-excitations of them and can propagate on the elastic substrate. We ourselves are wave-excitations, too, and because of this, the world appears Poincar\'{e} invariant to us, without a preferred rest system. 

In the beginning, that means before the expansion started, the universe was created in a sudden so to say inflationary condensation process from an ultrahot tetron gas and afterwards was pushed to expansion by the condensation energy. Matter and dark matter particles were created as quasiparticle excitations of the tetrons gliding on the elastic substrate. Since the quasiparticles fulfill Lorentz covariant wave equations, they perceive the universe as a 3+1 dimensional spacetime continuum lacking a preferred rest system. Any type of mass/energy induces curvature on the spacetime continuum as determined by the Einstein equations. According to the tetron-model\cite{tetrons} our universe is embedded in a higher-dimensional space, and as an elastic substrate it can thus acquire the full (3+1)-dimensional Einstein curvature within this space, including the timely curvature related to expansion. 

Since the substrate is assumed to be elastic, in (\ref{eq551x}) a harmonic ansatz $\ddot{L}=...-\omega^2 (L-L_s)$ seems reasonable, cf. the discussion after eq. (\ref{ab1}), in connection with (\ref{eq553}) and at the end of this work. Within such a picture, in an expanding universe, L and E vary with time (and so will h and G as well as all particle physics constants), and the next step is to make the most straightforward ans\"atze for these variations. 

As already noted, nowadays the universe is expanding towards an equilibrium corresponding to an average bond length $L_s$, i.e. it is carrying out an extremely low frequency breathing vibration $\omega \ll 10^{-10} yrs^{-1}$ around $L_s$, and all the tetron bonds on average are vibrating in accord with the universe. This draws a connection between the smallest and the largest scales of the universe, between the tetron binding structure at Planck length and the size of the universe as a whole. Namely, it implies that for all times t the bond length L(t) and the FLRW expansion parameter a(t) are proportional
%This is because within the breathing picture, the variation of the bond length L(t) reflects the general expansion as described by the FLRW expansion parameter a(t), in the sense that they are always proportional. In other words, the equivalence between (\ref{eq551x}) and the modified FLRW ansatz relies on the simple assumption that on the average the bond length between 2 tetrons is always proportional to the scale parameter, i.e. $a\sim L$ or equivalently 
\begin{eqnarray}
\frac{a(t)}{a_0}=\frac{L(t)}{L_0}
\label{eqttt553}
\end{eqnarray}
% or $a(t_1)/a(t_2)=L(t_1)/L(t_2)$ for all times $t_1$ and $t_2$.  
In other words, (\ref{eq551x}) is equivalent to a modified FLRW equation
\begin{eqnarray}
\ddot{a}=-\frac{4\pi}{3} G\rho L-\omega^2 (a-a_s) + \xcancel{\frac{\Lambda}{3} c^2 a}
\label{eq551xzz}
\end{eqnarray}
where $a_s$ represents the size of the universe at equilibrium $L_s$.

The FLRW scale a(t) describes changes at cosmological distances, while the bond length L(t) between tetrons is microscopic in origin. This is depicted schematically in Figure 1. The two parts of the figure represent the expanding universe at two times $t_2 > t_1$. a(t) corresponds to the full extension of the universe, while L(t) is the microscopic average spacing between two neighboring constituents. In Figure 1 the ratio $a(t_2):a(t_1)$ is given by 2. The same is true for the ratio $L(t_2):L(t_1)$, because for a breathing vibration the average bond length between the points grows in the same way as the system as a whole. 
%This simple consideration is at the heart of (\ref{eqttt553}) and also (through the FLRW equation) of (\ref{eq551x}). 

The first term in (\ref{eq551x}) arises from the general relativistic deceleration of the universe through its matter content $\rho$, while the second term accounts for the dark energy phenomenon, however, not quite in the usual form of a cosmological constant (crossed out, because not utilized in this part of the paper), but of a harmonic force $-\omega^2(L-L_s)$, that expands the elastic substrate towards an equilibrium value $L_s$ of the bond length L. 

Eq. (\ref{eq551x}) tells us that linear forces are acting, one induced by (ordinary and dark) matter and driving the system towards $L=0$, the other induced by the ('dark energy') tetron binding and driving it towards the equilibrium binding distance $L_s$. Presently we are in the region $L_0 < L_s$, so that $-\omega^2(L-L_s)$ really is an expanding force. The value of $\omega$ can be determined from a fit to dark energy measurements.

In the course of time, i.e. with increasing L,  the matter force becomes smaller because the matter density dilutes according to $\rho=\rho_0 L_0^3/L^3$. This is a well known effect and makes the first term on the RHS of (\ref{eq551x}) behave like $\sim 1/L$ instead of $\sim L$. 

The differential equation (\ref{eq551x}) can be solved using initial values 
\begin{eqnarray} 
L(0)=L_0 \qquad \quad
\dot L(0)=H_0L_0
\label{t3m3num}
\end{eqnarray}
where $L_0$ is the (present day) value of the Planck length and $H_0$ the Hubble constant (=present day value of the Hubble parameter $H(t)=\dot a /a=\dot L /L$). The solution will be given later in (\ref{eq558}). 
%(The qualitative solution for L(t) looks like in figure 1.)

From the initial conditions (\ref{t3m3num}) it is immediately clear that $\omega$ is naturally of the order of $H_0$. In section IV this will be roughly confirmed by fitting with observations. $\omega$ and H(t) are extremely small frequencies corresponding to an approximately harmonic movement of the universe as a whole and a priori have little to do with the Planck frequency $1/T_0$ which is the local response frequency of a single tetron in the elastic substrate. 
\begin{eqnarray}
%H_0=2.2\quad 10^{-18} \frac{1}{s}=1.18 \frac{10^{-61}}{T_0}
H_0\, T_0\approx 1.18 \times 10^{-61}
\label{eq554}
\end{eqnarray}
So seemingly, there are 2 very different fundamental scales in the universe: one is the single tetron binding energy/Planck energy E and the other is the collective dark energy of the universe as a whole, which drives it to its equilibrium value. 

However, due to the homogeneity of the elastic substrate, the time behavior of the microscopic single tetron binding energy E(t) and that of the cosmological dark energy turn out to be related. The important point is that the universe is in a breathing mode and its {\bf collective drive is just a reflection of the \bf microscopic tetron binding energy having a minimum at bond length $L_s$.} This becomes even clearer within the simple spring model described in appendix A, where one can relate $\omega$, the frequency of the universe introduced in (\ref{eq551x}), to the Planck time $T(t)$ via 
\begin{eqnarray}
\omega T(t_s) = \frac{1}{N}
\label{eq111554}
\end{eqnarray}
to be compared to (\ref{eq554}). N is the number of springs connected in a series from one end of the universe to the other. While the natural frequency of a single tetron spring is the Planck frequency $f(t)=1/T(t)$, when bound to form the universe, the series connection of N springs vibrates with a much lower frequency $\omega = f(t_s)/N$. 
%***(Frage: H(t)  ist doch gleich apunkt/a. Muss darueber nicht H0 mit $\omega$ verknuepft sein? omega haengt mehr mit azweipunkt zusammen. Nur azweipunkt sorgt fuer eine kruemmung. Insofern muss hubble auch mit azweipunkt zusammenhaengen und tut es in flrw auch!!!!)  

%Therefore, although the values of E and H are vastly different, the time dependencies E(t) and H(t) are connected. See eq. (\ref{eqttt5531}) later. 
It may be noted that within the simple spring model of appendix A, N is the constant of proportionality implicit in eq. (\ref{eqttt553}): 
\begin{eqnarray}
\frac{a(t)}{L(t)}=N=\frac{a_0}{L_0}
\label{eqttt573}
\end{eqnarray}

\begin{center}
{\bf III. Quantities-of-Motion Equation}
\end{center}

If one thinks it over, a time dependent L(t) has long been observed, namely in the form of the cosmolgical redshift. Usually this time dependence is not put into L, G or h, as in eqs. (\ref{tmeins}) and (\ref{tmzwei}), but into the redshifted photon frequency f and the expansion paramter a. This is possible, because these quantities always appear in products h*f and G*a, respectively. So one can choose whether to absorb the time dependence of L in h and G or in f and a. The conventional choice is to keep G and h constant. We shall follow this choice - as far as the variation of the Measure-of-Space equation is concerned.
%stimmt überhaupt, dass das Produkt G*a auftritt????? in der linearen Form von flrw schon!!!!
%außerdem: das ist gar nicht entscheidend, sondern dass a das gesamte L(t) auffängt

From this point of view, the ansatz of a time dependent L(t) is not so much new [apart from the modified cosmological constant approach to dark energy with $-\omega^2(L-L_s)$ instead of a $\Lambda$-term]. 

As for the time dependence of the Planck energy E, the situation is different, i.e. there will be something new: 

E can be interpreted as the binding energy among the constituents of the elastic substrate which is our universe. Not too far away from the equilibrium at $L=L_s$ it can be expanded in a power series of $(L-L_s)^2$ 
\begin{eqnarray}
E(L)=C + D(L-L_s)^2 + O(L-L_s)^4            
\label{ab1}
\end{eqnarray}
where the terms of order $(L-L_s)^4$ and higher will be neglected in the following. This corresponds to approximating the curve in Figure 2 by a parabola in the neighborhood of $L_s$, which should be a reasonable approximation not too far away from the minimum at $L=L_s$.

%(aber die planckenergie hat mit der planckfrequenz eine viel groessere frequenz als die kosmischen frequenzen omega die ich verwende - naja, es ist eben ein hoher proportionalitaetsfaktor dazwischen)

\begin{figure}
\begin{center}
\epsfig{file=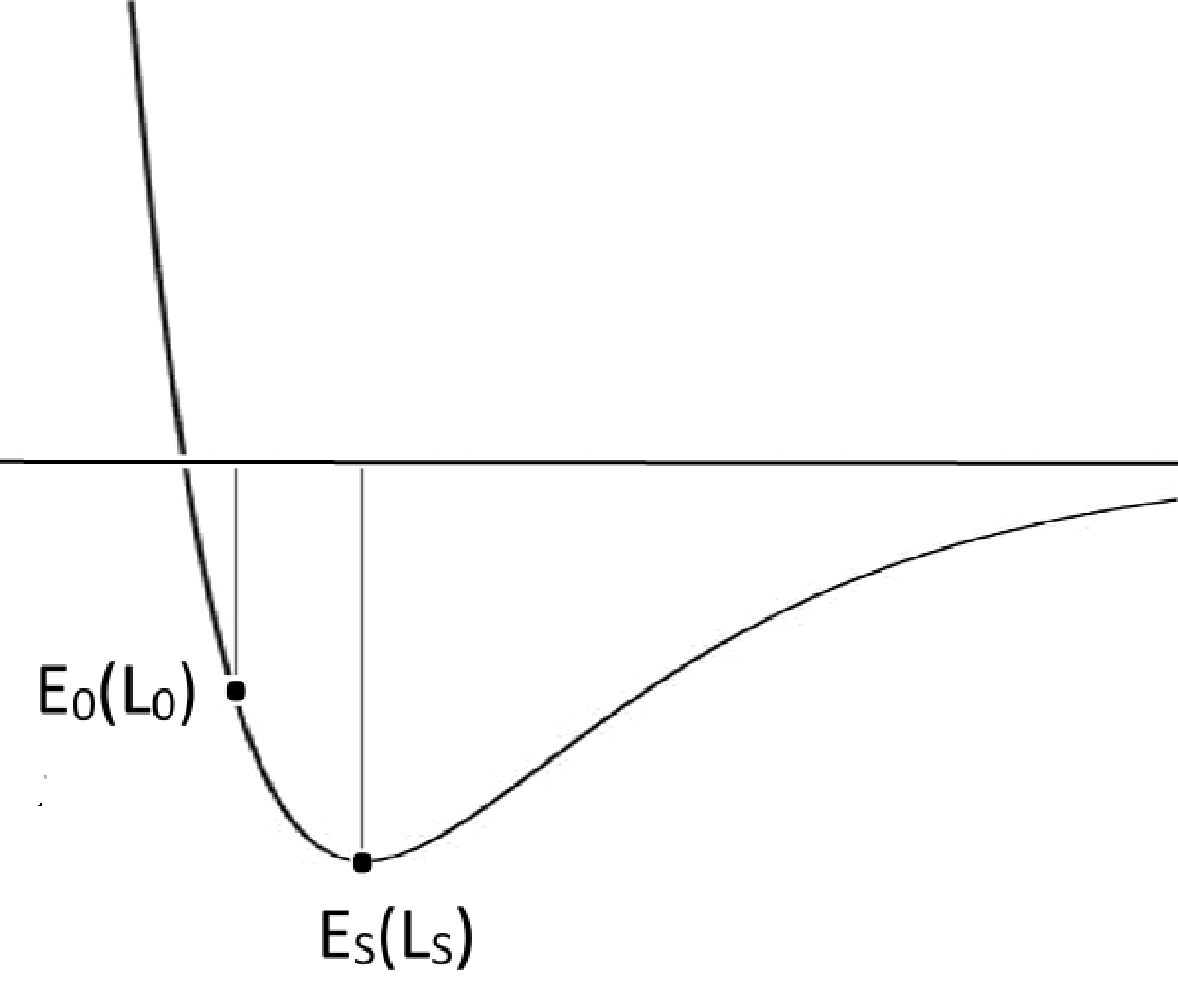,height=6.0cm}
%\bigskip
\caption{The binding energy E of 2 constituents as a function of their bond length L. At present t=0 one has the Planck energy $E_0$ and the Planck length $L_0$. The expansion of the universe through dark energy corresponds to the elastic bonds expanding towards equilibrium values $E_s$ and $L_s$. In the neighbourhood of $L_s$ the quadratic dependence E(L) of eq. (\ref{ab222}) is a good approximation.}
\nonumber
%\label{figac1} 
\end{center}
\end{figure}

The constants C and D can be determined from the conditions that $E(L_0)=E_0$ and $E(L_s)=E_s$. One obtains
\begin{eqnarray}
E(L)=E_s-(E_s-E_0) ( \frac{L-L_s}{L_0-L_s} )^2 =E_s [ 1-(1-\frac{E_0}{E_s})( \frac{1-L/L_s}{1-L_0/L_s} )^2 ]           
\label{ab222}
\end{eqnarray}
As will turn out, the energy difference $E_0-E_s$ triggers the harmonic dark energy term $\sim\omega^2$ in eq.(\ref{eq551x}), i.e. the accelerated expansion of the universe. 

The approximate behavior of E(L) is that of a parabola with minimum $E_s$. Together with the solution L(t) to (\ref{eq551x}) 
%(depicted in figure 2, and together with L(t) from figure 1) 
one deduces 
the time dependence E(t) as needed in eqs. (\ref{tmeins}) and (\ref{tmzwei}). 

Since we have absorbed the time dependence of L into the redshift definition, one only has to consider time dependencies of $\hbar$ and $G$ through $E(t)$, cf. eqs. (\ref{tmeins}) and (\ref{tmzwei}).
%Since we have absorbed the factors L(t) in eqs. (\ref{tmeins}) and (\ref{tmzwei}) into the redshift description, one only has to consider time dependencies according to 
%\begin{equation} 
%\hbar (t)\sim E(t) \qquad \qquad G(t)\sim 1/E(t)
%\label{tgg33b}
%\end{equation}
%or equivalently
%\begin{equation} 
%\hbar (t)=\hbar_0\frac{E(L)}{E(L_0)} \qquad \qquad G(t)=G_0\frac{E(L_0)}{E(L)}
%\label{tfrg33b}
%\end{equation}
These dependencies may be rewritten as
\begin{eqnarray} 
\hbar (t)/L&\sim & E(L)  \label{tfrg33b} \\
G(t)/L&\sim &1/E(L) \label{tfffff6} 
\end{eqnarray}
with $E(L)\equiv E(L(t))$ to be taken from (\ref{ab222}).

Considered as a binding energy, E(t) is negative, i.e. (\ref{tmeins}) and (\ref{tmzwei}) are better to be replaced by $\hbar (t) c=|E(t)| L(t)$ and $G(t)|E(t)|=c^4L(t)$ with $|E|=Mc^2$ being the Planck energy. 
%so one should better write $\hbar (t)\sim |E(t)|$ and $G(t)\sim 1/|E(t)|$. 
Since E(t) is negative and presently becomes more negative as it approaches its minimum value $E_s$, one concludes from eq. (\ref{tfrg33b}) that {\bf Plancks constant presently goes up with time, whereas the gravitational coupling is decreasing}.

At this point one may worry, whether a varying E has a problem with energy conservation. Actually, this question also arises in connection with the redshift, and is usually answered by saying that energy 'goes into the metric'. Interpreting the universe as an elastic substrate one can reformulate this by stating that energy goes into the total binding energy of the universe. 

As will be discussed in connection with eq. (\ref{eqffc777}), one can attribute the $\omega$-term in (\ref{eq551x}) to a time dependent   
\begin{eqnarray}
W(L)=\frac{\omega_s^2}{2}-\frac{\omega^2}{2} ( \frac{L-L_s}{L_0} )^2 
=\frac{\omega_s^2}{2} [1-(1-\frac{\omega_0^2}{\omega_s^2})( \frac{1-L/L_s}{1-L_0/L_s})^2]
\label{eq553}
\end{eqnarray}
with 
\begin{eqnarray}
\omega^2= \frac{\omega_s^2-\omega_0^2}{(1-L_s/L_0)^2}
\label{eq1553}
\end{eqnarray}
%***(stimmt die w-Notation mit dem appendix zusammen? Irgendwie ist da ein L**2 zuviel aber stoert letztlich nicht)
Note the similarity between (\ref{ab222}) and (\ref{eq553}). 
Since the dark energy phenomenon is a smooth collective effect of all tetron binding energies E having a minimum at bond length $L_s$, i.e. the behaviour of W is a reflection of the tetron bond length driving towards its equilibrium value $L_s$ (=the point where the tetron binding energy E is having a minimum value E$_s$), the time evolution of W and E is absolutely parallel, in an analogous way as the time evolution of a(t) is parallel to that of L(t). In other words, $E(L)\sim W(L)$ holds similarly as  $L(t)\sim a(t)$ for the cosmic scale factor a and the bond/Planck length L, cf. (\ref{eqttt553}) and Figure 1, and one comes up with
\begin{eqnarray}
\frac{W(L)}{W(L_0)}=\frac{E(L)}{E(L_0)}
\label{eqtut5531}
\end{eqnarray} 
The physical difference between W and E is that\\ 
-E is the microscopic tetron binding energy and is roughly of the order of the present day Planck energy to be measured in Joule.\\
-the $\omega$'s are frequencies of the universe as a whole and measured in Hertz, and they are of the order of the Hubble parameter.
%-must be, because $H_0$ enters the initial value of the differential equation (\ref{eq551x}).

A direct consequence of (\ref{eqtut5531}) is 
\begin{eqnarray}
\frac{\omega_0^2}{\omega_s^2}=\frac{E_0}{E_s}
\label{eqttt5531}
\end{eqnarray}

%The $\omega_s$ term is the energy of the equilibrium elastic substrate state, 
%and for $L=L_0$ we obtain the present day value of the dark energy as 
%\begin{eqnarray}
%\frac{\omega_0^2}{2}:=W(L_0)=\frac{\omega_s^2}{2}-\frac{\omega^2}{2} z_s^2 
%\label{eq553a}
%\end{eqnarray}
%where for later use I have introduced the abbreviation
%\begin{eqnarray}
%z_s =  \frac{L_s}{L_0} -1
%\label{eq553m}
%\end{eqnarray}
%Note further that here we are measuring energies in Hz$^2$, so as to be able to compare them to the %Hubble constant $H_0$. The frequency $\omega$ thus corresponds to the harmonic expansion and afterwards contraction of the universe.

\begin{center}
{\bf IV. Comparison with Astrophysical Data}
\end{center}

In the laboratory it is more or less impossible to observe time variations of G and h, because via (\ref{tm33b}) these quantities define our rulers for mass and energy. While the universe expands, the rulers will expand, too. 

In case of the redshift, astronomers were able to obtain relevant information on L(t) from observations of distant galaxies. In contrast, it seems difficult to measure the time variation (\ref{ab222}) of energy from such observations, because any process, which took place in the past in some distant galaxy, will do so with the energy/rulers relations valid at that time, and when the produced particles arrive on earth they will interact with the detectors with the energy/rulers relations valid now; so that the observer will see no difference between processes now and then. 

{\bf As a consequence, time variations of h and G will generally not be visible.} 

Hardly testable in particle processes, it turns out, however, that E(t) from eq. (\ref{ab222}) can be directly observed in dark energy measurements. Dark energy observations do not usually concern the very early universe, so that the parabolic approximation (\ref{ab222}) should be good enough\footnote{For considering time variations of h and G in the very early universe, an approximation of the form (\ref{ab222}) is not sufficient, because at small bond length L a typical binding energy is expected to be governed by a power behaviour of the form $E(L)\sim L^{-n}$.}.
%Furthermore, the matter contribution $\sim\rho$ in eq. (1) has to be taken into account, when solving for E(t), and will thus modify eq. (5).}
%They are in effect testing eq. (\ref{eq551x}), and E(t) in (\ref{ab222}) not only governs the $\omega$-term but according to (\ref{tfrg33b}) and (\ref{tgg33b}) also enters the G-term on the RHS of (\ref{eq551x}).

In order to check this idea with astrophysical data, we go over from L(t) to the redshift z defined by
\begin{eqnarray}
z(t) =  \frac{a}{a_0} -1=  \frac{L}{L_0} -1
\label{eq553m0}
\end{eqnarray}
    
%Since energy conservation in GR is a subtle business\cite{gibbs,weitere}, we start 
%with the 'force' equation
%\begin{eqnarray}
%\ddot{L}=-\frac{4\pi}{3} G\rho L-\omega^2 (L-L_s) + \color{green}{\frac{\Lambda}{3} c^2 a}
%\label{eq551}
%\end{eqnarray}
%already used above in eq.

The most precise measurement of the dark energy effect comes from the study of type-Ia supernovae in distant galaxies. I shall compare my redshift prediction to those data in a small-t approximation. This is justified because on cosmic scales the times involved are not too large. 

Under this condition, up to $O(t^4)$, the solution to (\ref{eq551x}) can be written as
\begin{eqnarray}
z=tH_0&+&\frac{t^2 H_0^2}{2}[-\frac{\Omega_M^0}{2}+
\begingroup
\uwave{\frac{\omega_0^2}{H_0^2} \frac{\frac{E_s}{E_P}-1}{\frac{L_s}{L_0}-1}} 
\endgroup
\begingroup
+\xcancel{\frac{\Lambda c^2}{3H_0^2}}
\endgroup
]
\nonumber\\
&+&\frac{t^3 H_0^3}{6}[\Omega_M^0 (1+
\begingroup
\uline{\frac{\frac{E_s}{E_P}-1}{\frac{L_s}{L_0}-1} }
\endgroup )
\begingroup
\uwave{-\frac{\omega_0^2}{H_0^2} \frac{\frac{E_s}{E_P}-1}{(\frac{L_s}{L_0}-1)^2}}
\endgroup
\begingroup
+\xcancel{\frac{\Lambda c^2}{3H_0^2}}
\endgroup
]
\label{eq558}
\end{eqnarray}
The term indicated by underlines is the contribution from the time dependent Newton constant, the terms with underwaves come from the harmonic dark energy $\omega$ contribution, and the crossed out terms from a cosmological constant (the latter to be ignored in this chapter).
\begin{eqnarray}
\Omega_M^0 = \frac{4\pi}{3} \frac{G_0\rho_0}{H_0^2}
\label{eq558a}
\end{eqnarray}
is the present day density parameter of matter in the universe, frequently used in this type of analysis. In the dark energy interpretation with a cosmological constant it comes out as roughly 0.3, 
which is usually considered a reasonable value.

As for any parabola, hidden in the parabolic dark energy (\ref{ab222}) and (\ref{eq553}) are 3 parameters, which need to be determined from observations. They may be chosen as\\
(i) $\frac{E_s}{E_0}=\frac{\omega_s^2}{\omega_0^2} > 1$ = the ratio of the Planck energies resp dark energies at cosmic equilibrium and at present\\
%\begin{eqnarray}
%\frac{E_s}{E_0}=\frac{\omega_s^2}{\omega_0^2}=xxx
%\frac{1}{1-z_s^2\frac{\omega^2}{\omega_s^2}} 
%\label{eq555}
%\end{eqnarray}
(ii) $\frac{L_s}{L_0} > 1$ = the ratio of the tetron binding lengths at cosmic equilibrium and at present\\
(iii) $\frac{\omega_0^2}{H_0^2} =$ the ratio of the present dark energy over the present value of the Hubble constant.\\
Since there are more parameters than in the ansatz of a cosmological constant, the observations will only give relations between i, ii and iii. Furthermore, an estimate for $\Omega_M^0$ has to be taken from other sources. Nevertheless, our next aim is to see what the observations allow to say.

A fit to the redshifts of supernovae yields\cite{perl} 
\begin{eqnarray}
z=tH_0+\frac{t^2 H_0^2}{2} (1.00\pm 0.05) +\frac{t^3 H_0^3}{6} (0.54\pm 0.05)
\label{eq558b}
\end{eqnarray}
Comparing with (\ref{eq558}) one finds that it is easy to accommodate the data with the help of the quantities i, ii and iii. For example, choosing $\Omega_M^0=0.3$ and\\
-$L_s=10L_0$ one obtains $E_s=1.34E_0$ and $\omega_0^2=3.4H_0^2$\\ 
-$L_s=2L_0$ one obtains $E_s=5.6E_0$ and $\omega_0^2=0.25H_0^2$
%In the former case the present day frequency of the universe is larger than the Hubble constant, in the latter case it is smaller. 

At first sight, the fact that data can be fitted this way so easily, seems to be a big surprise. After all, we are fitting numbers which usually are explained with an expontential increase due to a cosmological constant. The essential feature here is the contribution from the time variation of Newton's constant (underlined) which in combination with the harmonic dark energy contribution (underwaved) leads to an agreement with observations. The point is that since G(t) is going down with time, the retarding effect of (ordinary and dark) matter becomes smaller, and no exponential increase of the dark energy term as in the cosmological constant approach is needed to explain the supernova data. 

In other words, although the harmonic force ansatz corresponds to a more moderate re-acceleration of the universe than the cosmological constant term, this is compensated by the time variation of energy as a whole which affects Newton's constant.

%The new term arising from Gpunkt implies a relative enhancement of the 
%dark energy component as compared to the (ordinary and dark) matter component. 
%One may interpret this in either of 2 ways:\\
%-the gravitational force among matter becomes weaker 
%in comparison to the dark energy contribution\\
%-it is the dark energy itself, which is responsible for 
%the effective decrease in Newton's constant. 

\begin{center}
{\bf V. Cosmic Time Dependence of Particle Physics Parameters}
\end{center}

The discussion will now be extended to the 'constants', which describe the particle physics interactions. All parameters of the Standard Model (SM) of particle physics will be considered, i.e.\\ 
-the 3 dimensionless gauge couplings: the weak and electromagnetic fine structure constants $\alpha_{weak}$ and $\alpha$ together with the QCD scale parameter $\Lambda_{QCD}$.\\
-2 further parameters of the electroweak Standard Model, needed to describe the Higgs potential. These will be chosen to be the Higgs mass $m_H$ and the vacuum expectation value $v$ of the Higgs field.\\ 
%-the 2 parameters of the Higgs potential: the Higgs mass $m_H$ and the vacuum expectation value $v$ of the Higgs field. Note that using $v$ is equivalent to using the Fermi coupling $G_F=1/[\sqrt{2}v^2]$, and the quartic Higgs coupling is given by $\lambda=m_H^2/v^2$.\\
-the Yukawa couplings, which are all proportional to $v$.

Except for $\alpha_{weak}$ and $\alpha$, all these parameters have dimension of energy. If one looks at the definition of the fine structure constant
\begin{eqnarray}
\alpha = \frac{e^2}{4\pi\epsilon_0\hbar c}
\label{eqffc1}
\end{eqnarray}
it is the only dimensionless combination which can be built from the quantities $e^2/\epsilon_0$, h and c. As dimensionless, it is independent of the choice of rulers for time, length and energy\footnote{The dependence on the Planck/tetron binding energy E is not to be confused with the Wilsonian running of coupling constants, i.e. the dependence of $\alpha$ on the energies of particles in a scattering process.}. This is good news, because in looking for a cosmic time dependence of $\alpha$ one circumvents all the problems usually encountered in determining the time dependence of dimensionful quantities like E(t). The {\it bad news} in considering ratios like $\alpha$ is that most effects tend to drop out between numerator and denominator (see later). 
%(Note that epsilon0, the eleti perm is a factor which translates the electric cause into its mechanical effect.)

An interesting point is that although $\alpha$ itself is not an energy, it can be written as a ratio of forces or energies. Namely one can rewrite (\ref{eqffc1}) as
\begin{eqnarray}
\alpha =     \frac{e^2}{4\pi\epsilon_0 r^{(2)}} 
/ \frac{G_0M_0^2}{r^{(2)}}  
\label{eqffc2}
\end{eqnarray}
i.e. as the ratio of the electrostatic Coulomb (force) energy and the gravitational (force) energy of 2 point particles with elementary charge e and Planck mass $M_0$ at an arbitrary distance r. 

From this point of view the gravitational force is by no means weak as compared to the electric force, but - for such tetron-like test particles - is 137 times stronger!\footnote{Within the tetron model our universe is a 3-dimensional monolayer expanding within a 6- or 7-dimensional space, as described in Fig. 3 and Appendix B. It is actually possible to heuristically relate the value of $\alpha$ to the length scale $L_L$ which represents the thickness of the monolayer. To this end and using (\ref{eqffc3}) one may rewrite $\alpha$ as
\begin{eqnarray}
\alpha =     \frac{L_L E_L}{L_0 E_0}  
\label{eqfxfc2}
\end{eqnarray}
where today's Planck energy $E_0$ corresponds to the binding energy of a tetron along physical space and the new energy scale $E_L$ to the binding energy along the extra dimensions. However, since the positions of all tetrons within the monolayer are equivalent, their binding energies within the monolayer substrate are identical, i.e. $E_L=E_0$. This leads to the interesting result that the famous value $1:137$ of $\alpha$ may be roughly interpreted as the ratio $L_L:L_0$ of the magnitude of 1 tetrahedron within the extra dimensions to the distance of 2 of them in physical space, cf. Fig. 3. Furthermore, based on this consideration one may define an electromagnetic mass equicalent $M_L$ which amounts to
\begin{eqnarray}
M_L = \frac{M_0}{\alpha^2} \approx 0.44 \, g  
\label{eqfyfc2}
\end{eqnarray}
Note, $M_L$ is independent of any type of particle like electron, proton etc and allows to rewrite $\alpha$ as
\begin{eqnarray}
\alpha =     \frac{G_L M_L^2}{G_0 M_0^2}  
\label{eqffc2889}
\end{eqnarray}
with $G_L$ taking the role of an electromagnetic `Newton constant'. More details about the correspondence between the Planckian and the L-quantities can be found in Table 1 in Appendix B.\label{foo33}}

The key relation here is 
\begin{eqnarray}
\hbar c = G_0M_0^2 = L_0 E_0
\label{eqffc3}
\end{eqnarray}
which follows from (\ref{tm33bx}).
 
Defining $Q^2=e^2/[4\pi\epsilon_0]$ and introducing time dependencies, one has
\begin{eqnarray}
\alpha (t)= \frac{Q^2(t)/L(t)}{E(t)}  
\label{eqffc4}
\end{eqnarray}
where $Q^2$ comprises the electromagnetic effect in a measurement-system independent way. Obviously, Q$^2$ has the dimension of length$\times$energy. Since measurements and astrophysical observations show almost no time variation of $\alpha$, the time dependence of $Q^2/L$ must be the same as that of E(t) to a very good approximation. 

Referring once again to the tetron model, this has to do with the fact, that the time dependence of Q$^2$ is determined by that of the binding energy E(t)\cite{tetrons}, so that any time dependence of $\alpha$ drops out between numerator and denominator in (\ref{eqffc4}).
%\footnote{Another idea is that $\alpha$ reflects the ratio of internal and physical extensions, i.e. $R_0/L_0$ where $R_0$ is the diameter of an internal tetrahedron. If one then assumes that the tetrahedron expands at the same rate as ordinary space, $\alpha$ remains constant in cosmic time.} 

To understand this point in detail, one should note that the tetron model is more than a microscopic theory for the cosmic elastic substrate. The tetrons actually appear in the form of tetrahedrons which extend into a 3-dimensional internal space and whose excitations can be shown to represent the complete 3-family quark and lepton spectrum. For details see \cite{tetrons,lhiggs} and Appendix B. 

The internal interactions among tetrons are typical quantum interactions in the sense that one always has 'exchange' energies in addition to 'direct' energies, simply because for 2 (or more) identical particles - tetrons in this case - with single wave functions $f_1$ and $f_2$ their total wave functions are either symmetric or antisymmetric of the form $f_1(x_1)f_2(x_2)\pm f_1(x_2)f_2(x_1)$. Correspondingly, the relevant 2-point function of the tetron Hamiltonian can be described as the sum of the Planck(=binding) energy E(t) and a function J(t) usually called the exchange energy. In the present case it may be called 'internal exchange energy' because it arises as an integral including the extra dimensions.
\begin{eqnarray} 
E=\int d^6 x_1 \int d^6 x_2 f_1(x_1)f_2(x_2) V(1-2) f_1(x_1)f_2(x_2) \label{tmee}\\
J=\int d^6 x_1 \int d^6 x_2 f_1(x_1)f_2(x_2) V(1-2) f_1(x_2)f_2(x_1)  \label{tmjj} 
\end{eqnarray}
where the integrals are actually 6-dimensional, because they extend over both internal and physical space. V(1-2) is the potential between 2 tetrons with wave functions $f_1$ and $f_2$.\footnote{If one looks into the details of the tetron model\cite{tetrons}, the situation is a bit more complicated than described here. %First of all, $f_1$ and $f_2$ are the wave functions of tetron-antitetron {\it pairs}, and V(1-2) is the potential between the two pairs 1 and 2 sitting on neighboring corners of a tetrahedron. Secondly, 
Namely, to really calculate E and J from the 6-dimensional integrals one has to take the configuration of two adjacent tetrahedrons with at least 8 tetrons into account. Furthermore, there are actually two types of exchange integrals, one corresponding to the {\it inter}-tetrahedral interactions, which gives rise to the Fermi scale and is responsible for the large masses $m_t$, $m_W$ and $m_H$ of order 100 GeV, and another one corresponding to the {\it inner}-tetrahedral interactions, which gives rise to the lighter fermion masses and the QCD scale.\label{f1}} 

In a 6-dimensional environment the Green's function of the Laplace operator is $r^{-4}$, instead of $r^{-1}$ in the 3-dimensional case. Therefore, the most promising choice seems to be
\begin{eqnarray} 
V(1-2)=\frac{N}{|x_1-x_2|^4}
\label{tmjj0} 
\end{eqnarray}
with some coupling constant N. A rough estimate of N can be obtained by equating V(1-2) at the Planck length to the Planck energy. This gives a value for the fundamental tetron coupling N:

\begin{eqnarray} 
\frac{N}{L_0^4} \approx E_0  \quad \Longrightarrow \quad N\approx 10^{-130}\frac{m^6 kg}{s^2}
\label{tmjj066} 
\end{eqnarray}

When trying to calculate E and J according to (\ref{tmee}) and (\ref{tmjj}), one naturally runs into the so-called hierarchy problem of physics. Namely the question, why the relevant energy scales of gravity ($E_0\approx 10^{19}$GeV) and of particle physics ($J_0=1-100$GeV) are so much different. In the framework of the tetron model, the question can be reformulated: why is the exchange energy J so much smaller than the direct energy E? 

Looking at (\ref{tmee}) and (\ref{tmjj}), one sees that $J\ll E$ can happen, if the tetron wave functions are strongly localized. In the extreme case of delta functions one even finds, that the exchange integral vanishes, while the direct integral attains the value (\ref{tmjj066}). Such an extreme localization is of course unnatural. In order to get $J\approx 10^{-17}E$, it is enough to demand that f(x) drops from its maximum value at $x=0$ by about a factor of 10 at $x=L_0$. This is because J is a multidimensional integral and to integrate the product $dx_2 f_2(x_2) V(1-2) f_1(x_2)$ will give a suppression factor of roughly $\sim 0.1$ for each of the 6 dimensions. Similarly for the $x_1$-integration.

Except for $\alpha$, which is dimensionless and constant, I will argue that {\bf J(t) gives a universal time dependence for all internal/particle interactions - in a similar way as does the Planck energy E(t) for the spacetime quantities of motion}. In other words, while the time-dependence of all dimensionful spacetime quantities is dictated by E(t), the time dependencies of dimensionful SM particle properties like $v$, $m_H$, $m_W$ and all quark and lepton masses can be described in terms J(t). 

To see how this works in detail, one should relate J to the electroweak symmetry breaking scale. This was already done in \cite{tetrons}, where it was shown that the critical energy of the electroweak phase transition is given by an exchange integral J of the form (\ref{tmjj}). This is because in the tetron model the electroweak phase transition corresponds to an alignment of the tetrahedrons in the extra dimensions (`internal space'), and the Curie energy of this phase transition is given by J. Since the critical energy of the electroweak phase transition is approximately given by the Higgs vev $v$, one has $v=J$. 
%or, equivalently 
%\begin{eqnarray} 
%G_F(t)=\frac{1}{\sqrt{2}J^2(t)}
%\label{yj066} 
%\end{eqnarray}

%(Fussnote weglassen dass J nicht die tatsachliche Temp beschreibt, bei der er phasenubergang stattgefunden hat, da damals L groesser war)

It is well known that all particle masses in the SM are proportional to $v$. Therefore, J enters all dimensionful parameters of the electroweak SM - the fermion masses, the Higgs vev and the masses of the weak gauge bosons - in a linear way. All these quantities are $\sim J(t)$. 
%die Curietemperatur ist tatsächlich im wesentlichen durch das Austauschintegral gegeben!!!!
%3/2*k*Tcurie=z*j(j+1)*J
%wobei j=Gesamtspin und z=Zahl der relevanten nächsten nachbarn, siehe sehrgut curie...pdf 

Just as for E, the time-dependence of J arises through the time variation of the bond length L(t), i.e. through the expansion of the universe. If one would calculate the integrals J and E (\ref{tmee}) and (\ref{tmjj}) as a function of L, then knowing L(t) according to (\ref{eq551x}), one could deduce from that the time-dependence of $v$, of the Higgs mass and of all other parameters, and compare it to present upper limits\cite{altsch,uzan}.
%\footnote{Note that the actual temperature at which the electroweak phase transition took place in the universe, corresponds to some former value of L...} 
%da die SSB irgendwann stattgefunden haben muss, könnte man meinen, es kann keine Zeitabh von GF bzw vev geben. Jedoch wird GFheute bzw vevheute von dem derzeitigen Wert der Austauschenergie bestimmt, nicht vom damaligen.

Unfortunately, the situation is not that simple. First of all, as mentioned in footnote 
\ref{f1}, the integrals are difficult to calculate. Secondly, in everything we do, in every experiment we undertake, we encounter the Planck energy E(t) as a ruler, whose time dependence influences our perception of dimensionful quantities like $v$, $G_F$, $m_W$ and so on. To say it plainly, {\bf the time dependence we can perceive is not that of J(t) but that of the ratio J(t)/E(t).}

This means: if we consider, for example, a matter particle with mass $m_0$ in the present epoch, our perception of the time development m(t) of $m_0$ does not follow\footnote{Note there is no problem with the principle of equivalence because the heavy mass and the inert mass are both developing with J(t).}
\begin{eqnarray}
m(t)=m_0\frac{J(t)}{J_0} 
\label{1eqffc38}
\end{eqnarray}
but
\begin{eqnarray}
m(t)=m_0\frac{J(t)/J_0}{E(t)/E_0} 
\label{2eqffc38}
\end{eqnarray}
In the hypothetical case, that J and E would have an identical time dependence, the time dependence of m or of other dimensionful SM parameters could never be measured. 

By analyzing the structure of the direct and the exchange integral E and J (\ref{tmee}) and (\ref{tmjj}) in some detail, one can indeed show, that their dependence on the bond length L is quite similar, both with an extremum at nearly the same value $L_s$. Making an ansatz for J(L) analogous to that for E(L) in (\ref{ab222})
\begin{eqnarray}
J(L)=J_s-(J_s-J_0) ( \frac{L-L_s}{L_0-L_s} )^2 =J_s [ 1-(1-\frac{J_0}{J_s})( \frac{1-L/L_s}{1-L_0/L_s} )^2            
\label{ab1222}
\end{eqnarray}
one sees that the crucial part is the ratio $J_0/J_s$. To the extent that the equality
\begin{eqnarray}
\frac{J_s}{J_0} = \frac{E_s}{E_0}
\label{eqffc38}
\end{eqnarray}
%where $E_0$ and $J_0$ are the values of E and J now and $E_s$ and $J_s$ at the equilibrium. In this case, the time dependence would drop out in the ratio J/E, and practically all parameters of the SM would be independent of cosmic time. 
holds, a time dependence of SM parameters cannot be measured.
%\footnote{Such a statement would be in accord with the present status of observations, which only give upper limits on time dependencies of SM parameters.} 
Conversely, any observed time dependence in a SM parameter can be traced back to a deviation from (\ref{eqffc38}).

The integrals J and E can be analysed on a qualitative level, and according to this analysis the relation (\ref{eqffc38}) is approximately true. On the other hand there is no a priori reason, why it should be exactly true. First of all, the integrals (\ref{tmee}) and (\ref{tmjj}) are definitely distinct. Secondly, particle physics interactions have to do with inner symmetries not contained in the energetic analysis of the elastic universe [governed by E(t)]. Therefore, although present observations only give upper limits on time dependencies of SM parameters, their cosmic time dependence at least in principle follows its own rule, given by J(t).

\begin{center}
{\bf VI. Discussion}
\end{center}

In this study a theory concerning the time dependence of all known fundamental physical parameters has been developed. It rests on the idea that dark energy is a harmonic rather than an exponential effect, which is furthermore related to the binding energy of the underlying constituents of the universe. As has been shown, one is led to a time-dependence of Newton's and Planck's constant. These effects, however, are usually impossible to measure - except in the dark energy itself and in certain paricle physics properties.

Furthermore\\
-microsopic (L) und cosmic (a) length scales are connected in a simple linear kind of way ('the universe expands in the same manner as the tetron bonds expand'), cf. eq. (\ref{eqttt553}) and Figure 1.\\ 
-In an analogous fashion, Planck energies E(t) and dark energies $\omega$ are linearly related ('the total dark energy of the universe increases proportional to the single tetron binding energy') via (\ref{eqtut5531}). 

Since the universe is rather cool by now and apparently expands in a rather homogeneous way, these assumptions are expected to be very good approximations. This expectation was substantiated in section IV by proving that it leads to agreement with present day dark energy observations. Thereby it has turned out that there is a significant contribution to the observed dark energy effect from the time variation of Newton's constant. Since G(t) is going down with time, the retarding effect of ordinary matter becomes smaller, and no exponential increase of the dark energy effect as in the cosmological constant approach is needed. 

The Planck energy $E_0$ and its time-dependent generalization E(t) play a central role in the considerations presented here, see (\ref{tmeins}), (\ref{tmzwei}) and (\ref{ab222}). Actually, $E_0$ has been used in this paper with 2 meanings:\\
-it represents the gravitational energy of the interaction of 2 matter particles with Planck mass $M_0$ at Planck distance $L_0$, i.e. $E_0=G_0M_0^2/L_0$.\\
-it describes the binding energy of 2 tetrons bound at distance $L_0$.
%\\Although the binding energy interpretation may be not strictly identical to $E_0$ in numbers, its dependence on L resp t can be assumed to be the same as that of $E_0$.

Concerning the fundamental parameters of particle physics, it was shown that they depend on cosmic time via the internal exchange function J, whose dependence on L is similar but not exactly the same as that of E. With the advent of higher precision observations, this effect may become observable. 

A remaining problem is the calculation of E(t) (Planck energy) and J(t) (internal exchange energy) from first principles, i.e. from fundamental tetron interactions. In principle, this can be done using eqs. (\ref{tmee}) and (\ref{tmjj}), as soon as the precise form of the fundamental tetron interactions is known.

Finally, I want to discuss the question of energy conservation in the present framework. First of all, it should be stressed that energy is conserved locally in the interplay of processes between tetrons and ordinary and dark matter, just as it is in GR in the interplay between the metric, the vacuum energy and ordinary and dark matter. 

The possibility that the observed acceleration of the Universe and the possible time variation of the fundamental constants are two manifestations of the same underlying dynamics ('micro and macro connection'), has already been considered in a somewhat different context by the authors of \cite{fs1,fs2,fs3}. In those papers a cosmic time dependent 
vacuum energy density
%cosmological constant - a 'running vacuum' - 
has been introduced, and it was nicely shown how energy conservation can be maintained by an interplay between matter and the 'running vacuum'.

There is some similarity of this approach to the present work, however the arguments of \cite{fs1,fs2,fs3} are not directly applicable here, because the conserved energy of the elastic tetron substrate cannot be given in a Lorentz invariant way. Actually, the conserved quantity, from which the vibrational $\omega^2$-term in eq. (\ref{eq551x}) can be derived is given by $\dot{L}^2+\omega^2 (L-L_s)^2$, or, using (\ref{eqttt553}), by
\begin{eqnarray}
\frac{1}{2}\dot{a}^2+\frac{\omega^2}{2} (a-a_s)^2 = const
\label{eqffc777}
\end{eqnarray}
for a matter free elastic universe. $a_s$ is the equilibrium value of $a(t)$, at which the tetron binding energy has a minimum, cf. Figure 2. 

The $\omega^2$-term is not a part of the Einstein theory. It represents the energy of the elastic substrate universe which vibrates with a 'breathing frequency' $\omega$ around $a=a_s$. Furthermore, this expression is valid only in the FLRW cosmic coordinate system and therefore coordinate dependent and the model cannot be presented in a Lorentz covariant form.

The fact that it cannot be made a Lorentz covariant part of the Einstein theory should not be considered as a surprise, because general relativity is a theory of local curvature induced by energy-momentum and does not know about the equilibrium of the underlying elastic substrate at $a=a_s$. In the elastic picture of the universe, the FLRW cosmic coordinate system is a preferred system, where the tetrons do not move except for their coherent elastic expansion as in Figure 1. Ordinary and dark matter particles are quasi-particle waves, gliding on the tetron substrate. Since they respect Lorentz covariant wave equations, their physics looks the same in all inertial coordinate systems. The elastic tetron background, on the other hand, should be described on its preferred FLRW 'rest system' Figure 1, and for this the variational principle for (\ref{eqffc777}) can be formulated.

\begin{center}
{\bf Appendix A: A simple Spring Model for the empty Universe}
\end{center}

In this appendix a simple harmonic spring model is used to exemplify the ideas presented in the main text, in particular the idea of the universe as an extremely low frequency oscillatory system. 

Among others, it will be shown that by taking precise measurements on the cosmic frequency $\omega$ (i.e. of the dark energy effect), the full size and total mass/energy of the universe can be inferred. To obtain a value for $\omega$, the retarding effects from ordinary and dark matter have first to be subtracted out, as indicated in eq. (\ref{eq558}), but once this has been done and a value of $\omega$ extracted, one can use this value to determine the full size (and not just the observable size) of the universe. 

By 'full size' is meant the size of the 3-dimensional elastic substrate which once has condensed and now is expanding within some higher dimensional space (of dimension $\geq$ 6). As a measure of the size of the universe I shall take a(t) from eq. (\ref{tmas33b}) to have dimension of length and assume $dr^2$ to be dimensionless. 

In the course of discussion it turns out that a non-vanishing cosmological constant $\Lambda$ is needed after all, in addition to the oscillatory term $\sim \omega^2 (a-a_s)$. $\Lambda$ was left out in the main text eq. (\ref{eq551x}) for the sake of clarity of the arguments, but will be included in this appendix.

Furthermore, the equation of state of the dark energy, i.e. of the invisible tetron background substrate, will be derived\cite{silva}. The relation between tetron density and pressure is a characteristic property of the bound tetron system. Actually, the elastic tetron substrate resembles a fluid with elastic bonds (albeit with an extreme Lame constant) among its constituents rather than an ordered solid, and so the fluid equation seems an appropriate way of description.

The reason why the universe behaves harmonic is because its tetrahedral constituents follow a harmonic elastic interaction, i.e. the binding energy among the internal tetrahedrons in the neighborhood of the minimum at $L_s$ can be approximated by a parabola, see Figure 2. 

While characteristic frequencies $\omega$ and $H_0$ of the universe are tiny, the frequency $f_s=f(t_s)$ of a single tetron spring is extremely large and given by
\begin{eqnarray}
f(t)= \frac{1}{T(t)}=\frac{c}{L(t)} 
\label{e5qd}
\end{eqnarray}
where $T(t)=L(t)/c$ is the time dependent Planck time eq. (\ref{tm33b}). 
%***(2pi weggelassen NB: $f0^2/fs^2=E0/Es$)

\begin{center}
-
\end{center}

Coming to the details of the model, it is assumed that at each point of the elastic universe each direction can be approximated as a serial connection of N harmonic springs which connect N+1 constituents (= internal tetrahedrons). These can be thought to lie on a straight line running from one end of the universe to the other. The spatial extension of the universe is then given by
\begin{eqnarray}
a(t)=N L(t)
\label{eqffcnn}
\end{eqnarray}
where $L(t)$ is the bond=spring length as given in eq. (\ref{tm33b}). Clearly, this includes the additional simplifying assumption that the expanding condensate has approximately the same extension in all directions. 

The mass $\mu$ of any one of the spring chains is given by the sum of all constituent rest masses $M_{rest}$ in the chain, i.e. 
\begin{eqnarray}
\mu=N M_{rest}
\label{eqffcmm}
\end{eqnarray}
where $M_{rest}$ is the rest mass of one internal tetrahedron. The value of $M_{rest}$ is unknown but should typically be of the order of the Planck mass. 

According to eq. (\ref{ab222}), the Planck energy $E(t)$ (and similarly the Planck mass $M(t)=E(t)/c^2$) is a sum of a constant energy $E_s$ plus a variable component, which vanishes at $L(t_s)=L_s$ 
\begin{eqnarray}
E(L)=E_s + \frac{d}{2} (L-L_s)^2 
%=M_{rest}c^2-E_{bind} + \frac{M_{rest}c^2}{2L_s^2} (L-L_s)^2
=M_{rest}c^2-E_{bind} + \frac{1}{2}M_{rest}c^2 (1-\frac{L}{L_s})^2
\label{ab222fu}
\end{eqnarray}
$E_s$ comprises the binding energy of 2 tetrahedrons at $L_s$ as well as their possible rest mass
\begin{eqnarray}
E_s = E_{rest}-E_{bind}= M_{rest}c^2-E_{bind}
\label{eqffc83}
\end{eqnarray}
%Note, in this picture eqs 8 and 17 are simple consequences of (\ref{eqffcnn}) and (\ref{eqffcmm}), respectively.
%(man koennte auch sehr grosses Mrest haben und trotzdem Bindungszustand, weil man auf jeden Fall in einem Potentialtopf lebt.)
It is to be noted that (\ref{ab222}) and (\ref{ab222fu}) do not meet the condition $E(L=\infty)=M_{rest}c^2$, so they hold only in the neighborhood of $L=L_s$. Note further that $M_{rest}$ could in principle be any value, even much larger than $E_{bind}$. To have a bound state one needs a local minimum of $E(L)$ at $L_s$ but not necessarily a negative $E(L_s)$, i.e. the condition
\begin{eqnarray}
E_{bind}>E_{rest}
\label{ff222fu}
\end{eqnarray}
is not compelling - although present dark energy data seem to give a hint this inequality is true and furthermore in the simple spring model it will be needed for consistency of energy conservation, cf. eq. (\ref{eqfdtt9}).
\begin{center}
-
\end{center}

The spring constant of a single spring is $d=M_{rest} f(t_s)^2$ with $f(t)$ from eq. (\ref{e5qd}). 
The basic reason why the breathing frequency $\omega$ of the universe is so small whereas the fundamental frequency $f=1/T$ of its tetron constituents is so large, arises from the following fact: Consider one chain of strings stretching from one end of the universe to the other, each spring with a constant d. Then the serial connection of N springs is itself a harmonic oscillator with a much smaller combined spring constant
%\begin{eqnarray}
%\mu \ddot{a}(t) =-D \, [a(t)-a_s]
%\label{eqfdiff}
%\end{eqnarray}
\begin{eqnarray}
D=d/N 
\label{eqffcm12}
\end{eqnarray}
%In our model, The effective spring which extends from one end of the universe to the other is composed of many small springs connected in a series, and the effective spring constant of the universe can therefore be expressed from the constituent tetron spring constant via D. 
Note that the springs connected in parallel belong to different chains and do not contribute to the effective overall chain constant D. 
%(aber D enthält Faktor mue muss also gross sein???. Nein, D IST KLEIN klein aufgrund der folgenden Formel eqffcm23 gilt nämlich D/d=mue/M/N2=1/N was ok ist.) 

Using $D=\mu \omega^2$ one can express the (extremely small) frequency of the universe in terms of the (extremely large) Planck frequency $f_s=f(t_s)$:
\begin{eqnarray}
\omega^2 = \frac{D}{\mu}= \frac{d/N}{N M_{rest}}= \frac{1}{N^2} f_s^2 = \frac{c^2}{N^2L_s^2} 
\label{eqffcm23}
\end{eqnarray}

Thus, the full extension (\ref{eqffcnn}) of the universe at equilibrium can be given as 
\begin{eqnarray}
a_s = N L_s = \frac{c}{\omega}
\label{eqffcm25}
\end{eqnarray}
in a similar way as the observable(=Hubble) radius is given as $c/H_0$. The present size of the universe is somewhat smaller than $a_s$ and can be expressed as 
\begin{eqnarray}
a_0 = \frac{L_0}{L_s} \frac{c}{\omega}
\label{eqffcm261}
\end{eqnarray}
These equations show that a precise enough measurement of the dark energy effect (i.e. of $\omega$) can be used to determine the full size of the universe. 

%Just as the smallest and the largest scale are related in a simple linear kind of way ($a(t)=NL(t)$), so are the largest and the smallest frequency  ($f(t_s)=N\omega$). Using these 2 relations one can express the radius of the universe as $a(t)=f(t)L(t)/\omega(t)$. Thus its present day value is $a_0=2\pi L_0/T_0/\omega(t_0)$ with $\omega(t_0)=aus \omega$. 

\begin{center}
-
\end{center}

The spring model is also of use to better understand the FLRW theory within the general elasticity ansatz. Remember the essence of the FLRW model is contained in the following equations\\
(i) the Friedmann equation for the Hubble parameter $\dot{a}/a$ and\\ 
(ii) the `fluid equations' for the densities $\rho_{tet}$ and $\rho_{mat}$ of tetrons and (ordinary and dark) matter, respectively. Both, the tetron substrate and the matter content of the universe are assumed to be separate uniformly distributed perfect fluids with mass energy densities $\rho_{tet}(t)$ and $\rho_{mat}(t)$  and pressure $p_{tet}(t)$ and $p_{mat}(t)$.

In the present model one can write down an equation for the conservation of energy in a matter-free spatially flat tetron universe
%The Friedmann equation can actually be used to replace the 'frequency principle' eqs. (\ref{eqffc777}) and (\ref{eq553}) by a cosmological energy principle. This can easily be seen by writing a modified Friedmann equation for a matter free universe
\begin{eqnarray}
N^3 \frac{M_{rest}}{2}\dot{a}^2 +  3N^3\frac{M_{rest}c^2}{2} (1-\frac{a}{a_s})^2 
%-\xcancel{\frac{\mu}{6} \Lambda c^2a^2}  
= -N^3 E_s
%+\frac{1}{2}\mu_s kc^2
%\frac{4\pi}{3}G\rho_{mat} \mu_s a^2 + \frac{\mu_s}{6} \Lambda_{mat}c^2a^2
\label{eqfdtt1}
\end{eqnarray}
%(auch die spatial curvature has a contribution from matter and from tetron? Aber wird hier eh neglected)
the conserved energy being the kinetic plus potential energy of the system of $N^3$ masses $M_{rest}$ and $3N^3$ springs which make up the empty universe. The system is furthermore assumed to be in a breathing mode. Then, the difference in neighboring spring positions $x_{n+1}(t)-x_n(t)$ within one of the spring chains is n-independent and given by $L(t)-L_s$ for each spring n, and the velocity difference by $\dot{x}_{n+1}-\dot{x}_n=\dot{L}(t)$.

Comparing  (\ref{eqfdtt1}) to the corresponding matter-free FLRW equation, one finds two differences:\\
-there is no $\Lambda$-term in (\ref{eqfdtt1}). Any cosmological constant contribution from tetrons has been crossed out in  (\ref{eqfdtt1}), because for the matter free elastic tetron substrate the binding forces do not drive the universe with $\Lambda a^2$ to infinity, but with $\omega (a-a_s)^2$ to $a_s$. As will turn out later, in the presence of matter a cosmological constant reappears because the time dependent Newton constant (\ref{tmzwei}) implies a time dependent matter contribution to the cosmological constant\cite{fs1,fs2,fs3}.\\
-there is a nonvanishing constant contribution $\sim E_s$ on the rhs, where there is 0 in FLRW. This has to do with the fact that normally the Friedmann equation only counts contributions to the spacetime curvature, where a constant binding energy of the tetron constituents does not matter. The ordinary FLRW model does not know about the cosmic constituents and uses the freedom to put their background binding energy to zero. In other words, a constant (like the binding energy $E_s=E(L_s)$ at equilibrium $a=a_s$) does not contribute to any cosmic expansion. $E_s$ is the background tetron binding energy which leads to the flat elastic substrate being built from free tetrons, but does not at all contribute to its possible curvature or expansion.

%Comparing (\ref{eqfdtt1}) to a single bond energy (\ref{ab222fu}) between 2 tetrahedrons with rest mass $M_{rest}$ one obtains a formula for the tetron conribution to the cosmological constant
%\begin{eqnarray}
%\Lambda_{tet} \approx \frac{6}{a_s^2} (\frac{E_{bind}}{E_{rest}}-1)
%\label{cd1}
%\end{eqnarray}
%Depending on whether $E_{rest}$ is larger than $E_{bind}$ or vice versa, one obtains a positive or a negative value for $\Lambda_{tet}$. 

\begin{center}
-
\end{center}

So far we have considered an empty universe built up from an elastic substrate of bound tetrons (or, more precisely, of bound internal tetrahedrons). If, in addition, matter is present in the universe, the modified Friedmann equation becomes
\begin{eqnarray}
\frac{1}{2}\dot{a}^2 + \frac{3}{2} \omega^2 (a-a_s)^2 =
%-\frac{1}{6} \Lambda_{tet}c^2a^2  
%+\frac{1}{2}\mu_s kc^2
\frac{4\pi}{3c^2}G\rho_{mat} a^2 + \frac{1}{6} \Lambda_{mat}c^2a^2 
-\frac{E_s}{M_{rest}}
\label{eqfdtt9}
\end{eqnarray}
where $\rho_{mat}$ denotes the mass energy density of ordinary plus dark matter. To have consistency with aforegoing formulas, I have pulled out a factor of 3 in the definition of the cosmic frequency $\omega$.
%(da die rechte Seite fuer das ganze universum gilt, muss man die linke auch mit 3N2 multiplizieren??? Oder darf man einfach ueberall 3*N**2*mue weglassen?)
%***(Als wichtiger Hinweis: Es ist negativ, denn Im Sinne von eqffc83 steht auf der rechten Seite ein term -Es/Mrest=(Egesamt-Ebind)/Mrest, wobei Egesamt=1=c**2 das ist, was man erhalt, wenn man alles Andere auf die linke Seite schaufelt. Die G- und Lambda-Terme tragen negativ zu Egesamt bei und damit positiv zur Wurzel aus apunktquadrat. Egesamt-Ebind muss den negativen Beitrag des wTernes unter der Wurzel uberwiegen dh gross genug sein, damit die Wurzel aus apuunktquadrat gezogen werden kann!)

As before in (\ref{eqfdtt1}), the constant term $\sim E_s$ is not relevant when forming the time derivative of (\ref{eqfdtt9}) in order to obtain the equation for the curvature/acceleration $\ddot a$.  

Equation (\ref{eqfdtt9}) includes a cosmological constant contribution $\Lambda_{mat}$ due to ordinary and dark matter. This comes about as follows: 
%the total cosmological constant being given by 
%\begin{eqnarray}
%\Lambda =  \Lambda_{tet} +  \Lambda_{mat} 
%\label{efdt9}
%\end{eqnarray}
%While the $\Lambda_{tet}$-term corresponds to the bulk of the vacuum energy arising from the tetron subtrate, 
%As will be seen shortly, 
$\Lambda_{mat}$ measures a (small) classical matter vacuum energy. Since  in the present model the Newton constant $G=c^4 L(t)/E(L(t))$ is time dependent, $\Lambda_{mat}$ will turn out to be time dependent, too, an effect which for pedagogical reasons has been withhold in the main text. 

%The spatial curvature k represents a matter induced contribution to the total cosmic energy. Since tiny, it will be neglected for following discussion. 
%a cosmological constant contribution $\Lambda =  \Lambda_{tet} +  \Lambda_{mat}$ has been included both for tetrons and matter, for the following reasons:\\
%-The constant $\Lambda_{tet}$ is needed to accommodate the binding energy $E_s$ besides the quadratic term, cf. eq. 12. In other words, $E_s$ and $\Lambda_{tet}$ are related through
%\begin{eqnarray}
%-N [E_s-(E_s-E_0) ( \frac{L-L_s}{L_0-L_s} )^2]
%= \frac{\mu_s}{2} \omega^2 (a-a_s)^2 -\frac{\mu_s}{6} \Lambda_{tet}c^2a^2
%\label{eqfdtt2}
%\end{eqnarray}
%The overall minus sign on the lhs is due to the fact that binding energies E are counted negative.
%E0 hier ins spiel zu bringen ist mist- es geht auch einfacher. Dann sind die quadratischen Terme aus derPlanckfrquenz erhaltbar. Aber die KosmKonst muss mit a2=as2+O(a-as) gemacht werden und dann braucht man a-as term, der noch nicht mal in k drin ist! Oder variiert die Bindungsenergie mit a**2?\\
%-The matter contribution $\Lambda_{mat}$ will be needed for consistency of the continuity equation in case of a time dependent Newton constant. This will be made clearer after eq. imfolgende und should be included for a complete analysis in eq. 20.  

\begin{center}
-
\end{center}

In order to determine $\Lambda_{mat}$ and its time dependence, we now turn to the cosmological fluid equations. Both the expanding tetron background and the matter distribution will be approximated as separate fluids distributed homogeneously over the universe with energy densities $\rho_{tet}(t)$, $\rho_{mat}(t)$ and pressure $p_{tet}(t)$ and $p_{mat}(t)$. 

According to earlier discussions the tetronic constituents of the elastic universe behave more like a fluid than like a solid or a crystal. For this fluid the appropriate form of the fluid equation is the ordinary one
\begin{eqnarray}
\dot{\rho}_{tet} +3 (\rho_{tet} + p_{tet}) \frac{\dot{a}}{a} =0
\label{eqbi2}
\end{eqnarray}
because the spring coupling $d$ is constant and the time dependent Newton coupling not involved. In terms of a single tetrahedron, with physical volume $L^3$ around it, 
%***(das around it darf man nicht weglassen, da das eigentliche Volumen des Tetraeders intern ist) 
the density of the tetronic `dark energy' fluid is
\begin{eqnarray}
\rho_{tet}(t)=\frac{E}{V}=\frac{E_s}{L^3} + \frac{E_{rest}}{2L_s^3} (1-\frac{L(t)}{L_s}) ^2 
\label{eqfdtt2ax}
\end{eqnarray}
where $E_s=E(L_s)$ is the Planck energy at equilibrium and $M_{rest}=E_{rest}/c^2$ is the tetrahedral rest mass, cf. eq. (\ref{eqffc83}). 

Similarly, the dark energy pressure is obtained to be
\begin{eqnarray}
p_{tet}=-\frac{\partial E}{\partial V}=\frac{E_{rest}}{3L_s^3}(1-\frac{L}{L_s}) +O(L-L_s)^2
\label{eqfdtt2aa}
\end{eqnarray}
and it can be checked that (\ref{eqfdtt2ax}) and (\ref{eqfdtt2aa}) indeed fulfill eq. (\ref{eqbi2}). 

These equations correspond to an equation of state parameter 
\begin{eqnarray}
w=\frac{p_{tet}}{\rho_{tet}}=\frac{E_{rest}}{3E_s}(1-\frac{L}{L_s}) +O(L-L_s)^2
\label{eqfd2ee}
\end{eqnarray}
To compare this with various other dark energy equations of state suggested in the literature one may consult \cite{silva}.
%(Volumen als a**3 gewaehlt, aber waere es nicht besser als 4pi/3a3? Nicht wenn man die Tetraeder in einem kubischen Gitter anordnet. Ausserdem: Dann waere es klueger den omega2 Term von Anfang an mit einem Faktor 4pi/3 auszustatten)
%Remember that $E_s=E_{rest}-E_{bind}$ and that we are working throughout the paper in the neighborhoods of $L=L_s$ and $a=a_s$, respectively.

\begin{center}
-
\end{center}

For matter and dark matter the suitable form of the fluid equation can be derived from the Bianchi identity 
\begin{eqnarray}
T^\mu_{\nu;\mu}=0
\label{eqbi1}
\end{eqnarray}
for the energy-momentum tensor in general relativity. In case of a time dependent Newton and cosmological constant one has\cite{fs1,fs2,fs3}
\begin{eqnarray}
\frac{d}{dt} [G\rho_{mat} + \frac{c^4}{8\pi}\Lambda_{mat} ] + 3G(p_{mat} +\rho_{mat}) \frac{\dot{a}}{a} = 0
\label{eqfdtt2b}
\end{eqnarray}
For the late time cosmology under consideration, matter can be approximated in the standard way as uniformly distributed dust. Ordinary and dark matter should then fulfill the ordinary fluid equation
\begin{eqnarray}
\dot{\rho}_{mat}+3(p_{mat} +\rho_{mat}) \frac{\dot{a}}{a}=0
\label{eqfdtt2c}
\end{eqnarray}
by means of 
\begin{eqnarray}
p_{mat} &=&0  \\
\rho_{mat}(a)&=&\rho_{mat}(a_s) \frac{a_s^3}{a^3}
\label{tt25c}
\end{eqnarray}
Comparing (\ref{eqfdtt2b}) and (\ref{eqfdtt2c}) one concludes, that any imbalance coming from the time dependency of G must be cancelled by a time dependence of $\Lambda_{mat}$  according to\cite{fs1,fs2,fs3} 
\begin{eqnarray}
\dot{G} \rho_{mat} +  \frac{c^4}{8\pi}\dot\Lambda_{mat} = 0
\label{eqfdtt2d}
\end{eqnarray}

In the present approach all time dependencies arise only through $a(t)=NL(t)$. Therefore using 
\begin{eqnarray}
\frac{d}{dt} = \dot{a} \frac{d}{da}= \dot{L} \frac{d}{dL}
\label{eqfz1}
\end{eqnarray}
one can calculate $\Lambda_{mat}$ from the scale dependence of G eqs. (\ref{tmzwei}) and (\ref{ab222}) 
\begin{eqnarray}
\Lambda_{mat}(a)=\Lambda_{mat}(a_s)+8\pi  \int_a^{a_s} da \,\, \rho_{mat}(a) \frac{d}{da} \frac{L}{|E(L)|}
\label{eqfz2}
\end{eqnarray}
with $L=a/N$. In the quadratic approximation used throughout this paper, where one considers the neighborhood of $a=a_s$, one can carry out the integral in (\ref{eqfz2}) to obtain the a(t) dependence of $\Lambda_{mat}$: 
\begin{eqnarray}
\Lambda_{mat}(a)=\Lambda_{mat}(a_s)+\frac{8\pi\rho_{mat}(a_s)}{|E_s|} (L_s-L)
+O(L-L_s)^2
\label{eqfz3}
\end{eqnarray}
%The ratio $(L_s-L)/|E_s|$ can be interpreted as the ratio of the deviation $a-a_s=N(L-L_s)$ of the diameter of the universe from equilibrium $a_s$ over the energy $NE_s$ of the series of springs that stretches from one end of the universe to the other. 
Here, $\rho_{mat}(a_s)L_s^3/|E_s|$ is the ratio of average energy of matter within a Planck volume $L_s^3$ over one tetrahedral binding energy $E_s$. Since there are much more bound tetrons than matter particles in the universe, the cosmological constant due to (\ref{eqfz3}) is extremely small. This can be seen more explicitly by rewriting (\ref{eqfz3}) as
\begin{eqnarray}
\Lambda_{mat}(a)-\Lambda_{mat}(a_s)=8\pi\frac{\rho_{mat}(a_s)}{\rho_{tet}(a_s)}\frac{L_s-L}{L_s^3} \approx 8\pi\frac{\rho_{mat}(a_0)}{\rho_{tet}(a_0)}\frac{L_0-L}{L_0^3}
\label{eqfz41}
\end{eqnarray}
and using approximate values $L_0-L\approx -L_0$ and
\begin{eqnarray}
\rho_{mat}(a_0)/c^2 &=& 2.6 \, \, 10^{-27} \frac{kg}{m^3} \\
\rho_{tet}(a_0)/c^2 &=& \frac{M_0}{L_0^3} = 0.54 \, \, 10^{97} \frac{kg}{m^3}
\label{eqfz42}
\end{eqnarray}
where $M_0$ is the Planck mass eq. (\ref{tm3num}).
%***(wg lambdamat(a)-lambdamat(as)=10**(-130)*L0**(-2)=10**(-50)/meter**2 sind die Effekte wohl so gross wie die gemessene dark energy???)\\
%***(und damit kriegt man auch eine Abschaetzung der integrationskonstante Lambdamat(as)? Oder ist die integrationskonstante Lambdamat(as) die klassische kosmologische Konstante und als solche erst mal unbekannt? Wohl letzteres, und damit hat man das alte Problem zurück, wie groß das konstante Lambda ist)

\begin{center}
-
\end{center}

In the literature one often finds the argument that there should be enormous contributions to the cosmological constant from zero point fluctuations of the matter fields. This is `proven' by summing up the zero point energies $\hbar\omega_k/2$ of all existing fields over all modes with wave vectors k 
\begin{eqnarray}
\rho_{quant} = \frac{\hbar}{2} \int \frac{d^3 k}{(2\pi)^3} \omega_k  
\label{eqfz55}
\end{eqnarray}
and using the Planck scale as UV cutoff.
%$k_{cut}=2\pi/L_0$. 
This leads to an enormous number of the order of $\rho_{quant} \sim E_s/L_s^3$ - actually the same order as $\rho_{tet}$ eqs. (\ref{eqfdtt2ax}) and (\ref{eqfz42}) which is no wonder, because both the claimed quantum effects and the tetron properties are derived from the Planck scale. 

If true, such a huge contribution $\rho_{quant}$ to the cosmological constant would immediately tear the universe into pieces. This then is usually considered a fundamental unsolved problem and runs under the name `cosmological constant problem'. The suggested solutions often lead to a cascade of fine tuning problems.

In the tetron model the situation is much simpler. As discussed before, the Planck energy $E_0 \sim E_s$ contained in the neighborhood volume $L_0^3$ of one bound constituent is the binding energy of tetrons in an empty flat spacetime and does not affect the curvature(=accelerated expansion), because it is a constant contribution in the conservation of energy law (\ref{eqfdtt9}). Furthermore, quantum fluctuations do not exist by themselves at any point in the universe, but are an artefact  of the discreteness of the universe\cite{tetrons} and exist only where a matter field is gliding as a quasiparticle wave on the discrete tetron background. 

%For these reasons the cosmological constant does not get a significant contribution from quantum fluctuations.
\begin{center}
{\bf Appendix B: Basic Features of the Tetron Model}
\end{center}

The present article has been prepared within the framework of the tetron model\cite{tetrons,lhiggs}. Tetrons offer a microscopic understanding of gravitational as well as particle physics phenomena. In this appendix the basic assumptions of the model are listed as background information for the interested reader. A graphical representation of the essential ideas is shown in Fig. 3. 
\begin{figure}[p]
\begin{center}
\includegraphics[width=5.2in]{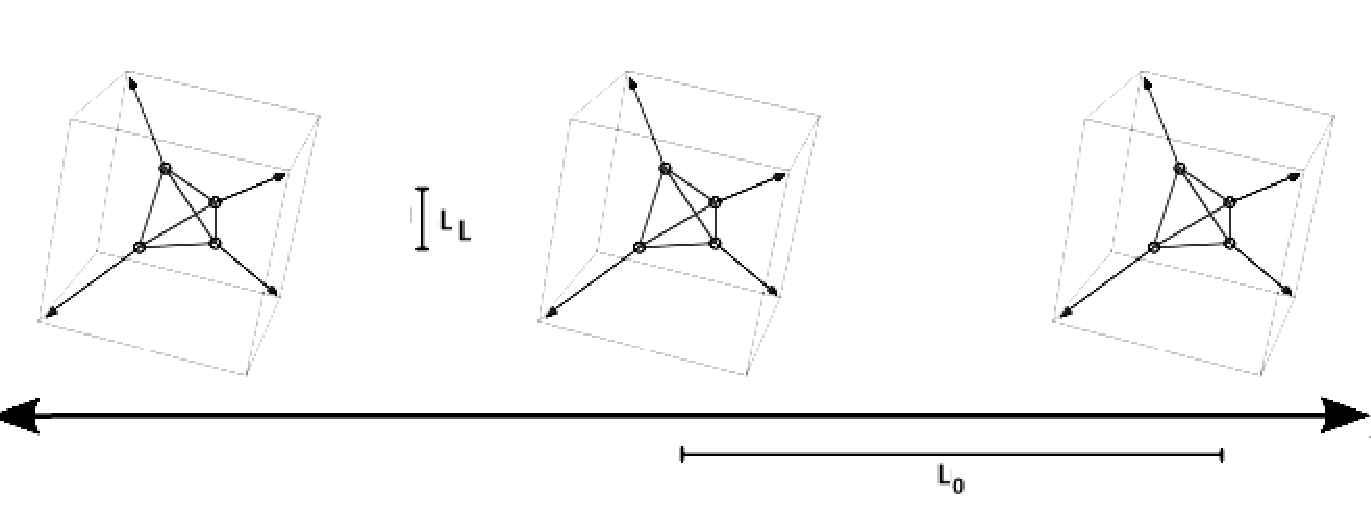}
\end{center}
\caption{The global ground state of the universe according to the tetron model, after the electroweak symmetry breaking has occurred, considered at Planck scale distances. The big black double arrow represents 3-dimensional physical space. $L_L$ is the magnitude of one tetrahedron within the 3 extra dimensions and $L_0$ the average distance between two neighboring tetrahedrons. The small arrows are the isospin vectors which are aligned in the ground state and whose eigen vibrations are responsible for the quark lepton spectrum. Note that the set of 4 arrows on each tetrahedron itself forms a tetrahedron and that each arrow stands in fact for 2 isospin vectors, namely for the ground states of $\vec Q_L$ and $\vec Q_R$ defined in (\ref{eq894bb}). The figure is a bit misleading, not only because the tetrahedrons do not extend into physical space, but also the relative magnitudes are not correctly drawn. Namely, while $L_L$ is of the order of the Planck length $L_0$, the extension of the tetrahedrons formed by the isospin vectors corresponds to the Fermi scale. 
%Actually there are 2 kinds of alignment: the alignment of neighboring coordinate tetrahedrons and the alignment of the tetrahedrons formed by the isospin vectors in neighboring tetrahedrons. 
While gravity can be attributed to the elasticity of the coordinate bonds, the phenomena of particle physics arise from the interactions between isospin vectors. 
Our universe thus is a 3-dimensional 'monolayer' of tetrahedrons (with thickness $L_L$) within a 6-dimensional space, each tetrahedron extending into the 3 extra dimensions. The monolayer ground state acts as a background on which quarks and leptons glide as quasiparticle excitations. It has the properties of a Lorentz ether and is thereby not in conflict with Michelson-Morley type of experiments.}
\label{aba:nneufig1}
\end{figure}

\vspace{-0.7cm}
\begin{center}
I.
\end{center}
%\centering{III.}
%\\ \raggedright
\vspace{-0.5cm}
-Our universe is a 3-dimensional elastic substrate, which once has condensed and now is expanding within some higher dimensional space, of dimension 6 or 7\footnote{As explained below under B.III, considering a 7+1 dimensional spacetime, instead of a 6+1 dimensional one, leads to a more consistent picture.}.
 
-The elastic substrate is built from invisible 'tetron' constituents with bond length about the Planck length and binding energy the Planck energy.

-Tetrons transform as the fundamental spinor(=octonion) representation 8 of SO(6,1) or SO(7,1). 

-All ordinary matter quarks and leptons are quasiparticle excitations of the tetrons gliding on the elastic substrate.
 
-Since the quasiparticles fulfill Lorentz covariant wave equations, they perceive the universe as a 3+1 dimensional spacetime continuum lacking a preferred rest system.
 
-Any type of mass/energy induces curvature on the spacetime continuum as determined by the Einstein equations.

-The Planckian quantities $L_0$, $T_0$ and $E_0$ relevant for space, time and energy are interpreted as the bond length, binding energy and hopping time within the substrate. The fine structure constant can be understood as the ratio of the length scales $L_L$ to $L_0$ depicted in Fig. 3.

\vspace{-0.7cm}
\begin{center}
II.
\end{center}
%\centering{III.}
%\\ \raggedright
\vspace{-0.5cm}
-The 24 known quarks and leptons arise as eigenmode excitations of a tetrahedral fiber structure, which is made up from 4 tetrons and extends into 3 extra ('internal') dimensions.

-With respect to the decomposition of $SO(6,1)\rightarrow SO(3,1)\times SO(3)$ into the 3-dimensional base space and the 3-dimensional internal space, a tetron possesses spin 1/2 and isospin 1/2. This corresponds to the fact that a tetron $\Psi$ decomposes into an isospin doublet $\Psi=(U,D)$ of two ordinary SO(3,1) Dirac fields U and D.

-Actually the tetrahedrons are formed by tetron isospin vectors of the form $\Psi^\dagger \vec\tau \Psi$, where $\vec\tau$ are the internal(=isospin) Pauli matrices. Due to the pseudovector property of the isospin vectors the tetrahedrons have a Shubnikov symmetry\cite{shub,borov,white}
\begin{eqnarray}
G_4:=A_4+CPT(S_4-A_4)
\label{eq8gs}
\end{eqnarray}
where $A_4 (S_4)$ is the (full) tetrahedral symmetry group and CPT the usual CPT operation except that P is the parity transformation in physical space only. Since the elements of $S_4-A_4$ contain an implicit factor of internal parity, the symmetry (\ref{eq8gs}) certifies CPT invariance of the local ground state in the full of $R^{6+1}$. The tetrahedral Shubnikov group $G_4$ is unbroken and holds down to the lowest energies.
 
-While the laws of gravity are due to the elastic properties of the substrate formed by the tetrahedrons, particle physics interactions take place within the internal fibers, with the characteristic interaction energy being the Fermi scale.
 
-Masses and mixings of quarks and leptons can be determined from a Heisenberg type of Hamiltonian which describes the interactions of any two neighboring tetron isospins and gives rise to 24 eigenmodes.
 
-3 of the 24 eigenmode excitations naturally have tiny masses, due to the conservation of total isospin(=internal angular momentum), and correspond to the 3 neutrino species.
 
-As a consequence of the free rotatability of the tetron isospins over each base point there is a local $SU(2)\times U(1)$ gauge symmetry. The corresponding gauge fields as well as the Higgs scalar can be constructed as combined excitations of isospin vectors from 2 neighboring tetrahedrons\cite{bodoeps}.
 
-Electroweak spontaneous symmetry breaking takes place when neighboring tetron isospins align (freeze out) at energies/temperatures below the Fermi scale. 
 
-Weak parity violation arises from the handedness of the internal tetrahedrons formed by isospins.
 \begin{center}
\begin{table}
\begin{center}
\begin{tabular}{ | l | l |}
\hline
Planckian quantities & L-quantities \\
\hline
$L_0$ & $L_L=\alpha L_0$ \\
$T_0$ & $T_L=T_0$ \\
$E_0$ & $E_L=E_0$ \\
\cline{1-2}
$M_0=E_0/c_0^2$ & $M_L=M_0/\alpha^2\approx 0.44\, g$ \\
\hline
\hline
$c_0=L_0/T_0$ & $c_L:=L_L/T_L=\alpha c_0$ \\
$h_0=E_0 T_0$ & $h_L=E_L T_L=h_0$ \\
$\kappa_0=L_0/E_0$ & $\kappa_L=L_L/E_L=\alpha \kappa_0$ \\
\cline{1-2}
$G_0=\kappa_0 c_0^4$ & $G_L=\kappa_L c_L^4=\alpha^5 G_0 \ll G_0$ \\
\hline
\end{tabular}
\caption{Confronting the roles of the length scales $L_0$ and $L_L$ defined in Fig. 3. The table is based on the discussion in Footnote \ref{foo33}, i.e. to the re-interpretation of the fine structure constant as $L_L/L_0$ and the associated definition of an electromagnetic mass equivalent $M_L=M_0/\alpha^2\approx 0.44\, g$. Note, the so defined $M_L$ is independent of any particle type like electron, proton etc, i.e. it is solely a property of the tetron substrate. Furthermore, the smallness of $G_L$ in comparison with $G_0$ is related to the smallness of the Fermiscale in comparison to the Planck scale. The point is that the exchange integrals determining Fermi scale and electroweak gauge couplings are much smaller\cite{tetrons} than the direct integrals responsible for the value of $G_0$.}
\end{center}
\end{table}
\end{center}
 
%(*) Tetrons transform as the spinor representation 4 of SO(6) resp 8 of SO(6,1).
%(**) Actually the tetrahedrons are formed by tetron isospin vectors. Due to the pseudovector property of the isospin vectors they have a (tetrahedral) Shubnikov symmetry.
%The Hamiltonian can be written in terms of the tetron isospin vectors, see the enclosed review talk.

\vspace{-1.0cm}
\begin{center}
III.
\end{center}
%\centering{III.}
%\\ \raggedright
\vspace{-0.5cm}
Here the case of a 7+1 dimensional spacetime is made explicit. In 7+1 dimensions there are two complex spinorial octonion representations $8_L$ and $8_R$\cite{ross,slansky}, one with left handed and the other with right handed chirality. When decomposing $SO(7,1)$ into the Lorentz symmetry $SO(3,1)$ of the base Minkowski spacetime and the internal $SO(4)$, one obtains\cite{slansky} 
\begin{eqnarray}
SO(7,1)&\rightarrow& SO(3,1)\times SU(2) \times SU(2) \nonumber \\
8_L &\rightarrow& ((1,2),(2_L,1))+((2,1),(1,2_R)) \nonumber \\
8_R &\rightarrow& ((1,2),(1,2_R))+((2,1),(2_L,1)) \nonumber \\
8_L \oplus 8_R &\rightarrow& ((1,2)+(2,1),(1,2_R)+(2_L,1))
\label{eq871}
\end{eqnarray}
where the covering group $SU(2)\times SU(2)$ of $SO(4)$ has been introduced. $(2,1)$ and $(1,2)$ denote left and right handed Weyl spinor representations of $SO(3,1)$ while the left and right handed $SO(4)$ spinors are denoted by $(2_L,1)$ and $(1,2_R)$. Note, the chiralities of all the doublets appearing in (\ref{eq871}) follow from the chiralities of the parent representations $8_L$ and $8_R$.

In order to obtain the necessary 24 internal (i.e. $SO(4)$) vibrational degrees of freedom, the decompositon of $8_L \oplus 8_R$ in the last line of (\ref{eq871}) should be considered. The corresponding tetron field represents a Dirac field $(1,2)+(2,1)$ in the $SO(3,1)$ base part, with 2 $SO(4)$ doublets $\Psi_L=(U_L,D_L)$ and $\Psi_R=(U_R,D_R)$ on top. It is important to note that here the indices L and R refer to the chiralities in the 4 extra dimensions. Accordingly, the isospin angular momentum vectors should be defined with respect to the two $SU(2)$ subgroups in $SO(4)$, and not using the $\gamma_5$ of the Lorentz group but $\gamma_5^{SO(4)}$ defined in $SO(4)$:
\begin{eqnarray}
\vec Q_L=\frac{1}{4}\Psi^\dagger (1-\gamma_5^{SO(4)})\vec \tau\Psi =\frac{1}{2}\Psi^\dagger_L \vec\tau \Psi_L
\quad 
\vec Q_R= \frac{1}{4}\Psi^\dagger (1+\gamma_5^{SO(4)})\vec \tau\Psi=\frac{1}{2}\Psi^\dagger_R \vec\tau \Psi_R 
\label{eq894bb}
\end{eqnarray} 
and similarly for the tetron densities $n_L$ and $n_R$. 
%The appearance of $\gamma_5$ has to do with the fact that the chiralities within the representations are intertwined, as implicit in (\ref{eq871}). 
%das folgende ist falsch da dann tensorprodukt 8x8 nötig werden: equivalent to considering 2 tetrons on each corner as in $SO(6,1)$. 

On each tetrahedral site there are thus 8 independent internal vibrators, namely $\vec Q_L$, $\vec Q_R$, $n_L$ and $n_R$. A tetrahedron with 4 sites then gives 32 vibrational eigenstates, 24 of which make up for the observed quarks and leptons and the remaining 8 density vibrations for a family of dark matter.
This point of view effectively is equivalent to having two $SO(6,1)$ tetrons on each tetrahedral site, cf. Fig. 5 in \cite{tetrons}.
%(dies ist besser, da 8 von so61 nur ReD,ReU,ImD,ImUx4=16 dof hat.)
%(in so61 muss man daher 2 tetronen an jeder Ecke annehmen. Nein, denn führt zu tensorprodukten. Auch wieder falsch, denn Tensorprodukte gibt es nur bei Bindungszuständen mit Punkten x ungleich y über dem Minkowskiraum)
%Working out the details it turns out that the dominant contribution to the Z-mass is from $V_{UU,UU }$, similarly to 6+1.

Further advantages of $SO(7,1)$ over $SO(6,1)$ are\\
-that wave functions obey Huygens' principle in 7+1 but not in 6+1 dimensions\cite{atiyah} and\\
-the fact that $G_4$ is a subgroup of only one of the 2 isospin $SU(2)$s may affect and simplify the derivation of weak parity violation given in \cite{tetrons}.

%-------------------------------------------------------------


\begin{thebibliography}{99}
\bibitem{dirac} P. A. M. Dirac, Nature 139, 323 (1937); Proc. Roy. Soc. A165, 199
(1938).
\bibitem{uzan} see the review by J.P. Uzan, Living Reviews in Relativity 14 (2011) 2.
\bibitem{duff1} M. J. Duff, L. B. Okun and G. Veneziano, JHEP 0203, 023 (2002).
\bibitem{moffat1} J. W. Moffat, Int. J. Mod, Phys. D2, 351 (1993).
\bibitem{barrow1} J. D. Barrow, Phys. Rev. Lett. 82 (1999) 884.
\bibitem{beken1} J. D. Bekenstein, J. D. Phys.Rev. D25 (1982) 1527. 
%\bibitem{beken2} Bekenstein, J. D. Phys.Rev. D66 (2002) 123514 gr-qc/0208081 
\bibitem{fs0} H. Fritzsch and J. Sola, Adv. High Energy Phys. 2014 (2014) 361587.
\bibitem{mag11} J. Magueijo, Rept. Prog. Phys. 66 (2003) 2025. 
\bibitem{webb1} J. K. Webb et al. Phys.Rev.Lett. 87 (2001) 091301.
%General Dynamics of Varying-Alpha Universes - Barrow, John D. et al. Phys.Rev. D88 (2013) 103513 arXiv:1307.6816 [gr-qc] 
%Spatial variations of fundamental constants - Barrow, John D. et al. Mon.Not.Roy.Astron.Soc. 322 (2001) 585 astro-ph/9904116 
\bibitem{belin} J. A. Belinchon et al., Int. J. Mod. Phys. D12 (2003) 1113.
\bibitem{carv} J. C. Carvalho et al. Phys. Rev. D46 (1992) 2404.
%Time Variation of the Fundamental 'Constants' and Kaluza-Klein Theories - Marciano, William J. Phys.Rev.Lett. 52 (1984) 489 BNL-33980 
%\bibitem{kolb} E. W. Kolb et al. Phys. Rev. D33 (1986) 869 FERMILAB-PUB-85-136-A 
\bibitem{take} T. Chiba, Prog. Theor. Phys. 126 (2011) 993.
%Constraining variations in the fine - structure constant, quark masses and the strong interaction - Murphy, Michael T. et al. Lect.Notes Phys. 648 (2004) 131-150 astro-ph/0310318 
\bibitem{calm} X. Calmet et al., Phys. Lett. B540 (2002) 173.
%Variations of alpha in space and time - Barrow, John D. et al. Phys.Rev. D66 (2002) 043515 astro-ph/0202129 
\bibitem{gard1} C. L. Gardner, Phys. Rev. D68 (2003) 043513.
\bibitem{mota} D. F. Mota et al. Phys.Lett. B581 (2004) 141.
\bibitem{park} D. Parkinson et al. Phys. Lett. B578 (2004) 235.
\bibitem{cope} E. J. Copeland, Phys. Rev. D69 (2004) 023501.
%Bouncing universes with varying constants - Barrow, John D. et al. Class.Quant.Grav. 21 (2004) 4289-4296 astro-ph/0406369 
%Shaw, Douglas J. et al. Phys.Rev. D73 (2006) 123505 gr-qc/0512022 
%Time Variation of Fine Structure Constant and Proton-Electron Mass Ratio with Quintessence - Lee, %Seokcheon Mod.Phys.Lett. A22 (2007) 2003-2011 astro-ph/0702063 [ASTRO-PH] 
%Varying alpha and the electroweak model - Kimberly, Dagny et al. Phys.Lett. B584 (2004) 8-15 hep-ph/0310030 
\bibitem{shaw1} D. J. Shaw et al., Phys. Rev. D71 (2005) 063525. 
\bibitem{tetrons} for a review see B. Lampe, arXiv:1505.03477 [hep-ph]. 
\bibitem{lhiggs} B. Lampe, Int. J. Mod. Phys. A30 (2015) 1550026.
\bibitem{fs1} H. Fritzsch and J. Sola, Class. Quant. Grav. 29 (2012) 215002.
\bibitem{fs2} H. Fritzsch and J. Sola, Mod. Phys. Lett. A, Vol. 30, No. 22 (2015) 1540034.
\bibitem{fs3} H. Fritzsch, J. Sola and Rafael C. Nunes, Eur.Phys.J. C77 (2017) 193.
%\bibitem{gibbs} see e.g. J. Baez, math.ucr.edu/home/baez/physics/Relativity/GR/energy.html or P. Gibbs, viXra:1305.0034 or S.M. Carroll, arXiv:gr-qc/9712019. 
\bibitem{perl} A.G. Riess et al., Astronomical Journal 116 (1998) 1009; S. Perlmutter et al., Astrophysical Journal 517 (1998) 565.
%\bibitem{lmass} B. Lampe, arXiv:1405.6604 [hep-ph], Int. J. Mod. Phys. A30 (2015)1550025.
%\bibitem{la4xz2} B. Lampe, arXiv:1201.2281 [hep-ph], Int. J. Theor. Phys. 51 (2012) 3073.
%\bibitem{llett} B. Lampe, arXiv:1212.0753 [hep-ph], Mod. Phys. Lett. A28 (2013) 135.
%\bibitem{lfound} B. Lampe, arXiv:0805.3762 [hep-ph], Found. Phys. 40 (2009) 573.
\bibitem{altsch} A. Ferrero and B. Altschul, arXiv:1008.4769v2 [hep-ph]
\bibitem{silva} R. Silva, R. S. Goncalves, J. S. Alcaniz and H. H. B. Silva, Astron. and Astrophys. A11 (2012) 537.
\bibitem{shub} A.P. Cracknell, Progr. Theor. Phys. 35 (1966) 196.
\bibitem{borov} A.S. Borovik-Romanov and H. Grimmer, International Tables for Crystallography D (2006) 105.
\bibitem{white} R.M. White, Quantum Theory of Magnetism, Springer Verlag, ISBN-10 3-540-65116-0.
\bibitem{bodoeps} B. Lampe, PoS(EPS-HEP2023), 449 (2024) 373.
\bibitem{ross} G. Ross, Grand Unified Theories, Cummings Publishing Company, 1984.
\bibitem{slansky} R. Slansky, Phys. Rept. 79 (1981) 1. 
\bibitem{atiyah}  M.F. Atiyah et al., Acta Mathematica 124 (1970) 109.
\end{thebibliography}
\end{document}